\documentclass[pre,superscriptaddress,tightenlines,showpacs,nofootinbib,notitlepage]{revtex4-1}
\usepackage{graphicx,amssymb,amsmath,epsf,bm}
\usepackage[normalem]{ulem}
\usepackage{stackrel}
\usepackage{color}
\usepackage{upgreek}

\usepackage{latexsym,amsmath,amsfonts,tabularx,amssymb,bm,empheq}
\usepackage{graphicx,epstopdf}
\usepackage{xspace}

\usepackage{hyperref}
\hypersetup{
  citecolor = {green},
  colorlinks = {true}, % false
  urlcolor = {blue} % magenta
}

\newcommand{\bea}{\begin{eqnarray}}
\newcommand{\eea}{\end{eqnarray}}

\def\s{\mathbf{s}}
\def\q{\mathbf{q}}

\def\x{\textbf{x}}
\def\y{\textbf{y}}
\def\h{\mathbf{h}} 
 
\def\bs{\boldsymbol{\sigma}}

\def\tcrhos{\breve\rho}
\def\trho{\tilde\rho}

\def\hbphi{\hat{\boldsymbol{\phi}}}

\newcommand{\hrho}{\hat{\rho}}
\newcommand{\crho}{\check{\rho}}
\newcommand{\clambda}{\check{\lambda}}
\newcommand{\bphi}{\bar{\phi}}
\newcommand{\hsigma}{\hat\sigma}
\newcommand{\csigma}{\check\sigma}
\newcommand{\hbpi}{\hat{\boldsymbol{\pi}}}

\def\uvtheta{\upvartheta}
\def\cF{\check{F}}
\def\cW{\check{W}}
 
 \def\bphi{\boldsymbol{\phi}}   
     
\newcommand{\DD}{\mathcal{D}}
\newcommand{\ZZ}{\mathcal{Z}}
\newcommand{\HH}{\mathcal{H}}
\def\Tr{{\rm Tr}}

\def\simge{\mathrel{%
   \rlap{\raise 0.511ex \hbox{$>$}}{\lower 0.511ex \hbox{$\sim$}}}}
\def\simle{\mathrel{
   \rlap{\raise 0.511ex \hbox{$<$}}{\lower 0.511ex \hbox{$\sim$}}}}

\def\simle{\mathrel{
   \rlap{\raise 0.511ex \hbox{$<$}}{\lower 0.511ex \hbox{$\sim$}}}}

\def\simge{\mathrel{%
    \rlap{\raise 0.511ex \hbox{$>$}}{\lower 0.511ex \hbox{$\sim$}}}}

\usepackage{subfigure}
\usepackage{color}
\usepackage[dvipsnames]{xcolor}
\usepackage{tabularx}

\begin{document}

\title{
Probability distributions of the order parameter of the $O(N)$ model
}

\author{A. Ran\c con}
\affiliation{
Univ. Lille, CNRS, UMR 8523 -- PhLAM -- Laboratoire de
Physique des Lasers Atomes et Mol\'ecules, F-59000 Lille, France
}
\affiliation{Institute of Physics, Bijeni\v{c}ka cesta 46, HR-10001 Zagreb, Croatia}
\author{B. Delamotte}
\affiliation{
Laboratoire de Physique Th\'eorique de la Mati\`ere Condens\'ee, UPMC,
CNRS UMR 7600, Sorbonne Universit\'e, 4, place Jussieu, 75252 Paris Cedex 05, France
}
\author{L. \v{S}aravanja}
\affiliation{Department  of  Physics,  University  of  Zagreb,  Bijeni\v{c}ka  c. 32,  10000  Zagreb,  Croatia}
\author{I. Balog}
\affiliation{Institute of Physics, Bijeni\v{c}ka cesta 46, HR-10001 Zagreb, Croatia}

\date{\today}

% \pacs{05.70.Ln, %Nonequilibrium and irreversible thermodynamics
% 05.50.+q,
% 02.50.-r,
% 64.60.Ht %Dynamic critical phenomena}
% }

\begin{abstract}
    We study the probability distribution function (PDF) of the order parameter of the three-dimensional $O(N)$ model at criticality using the functional renormalisation group. For this purpose, we generalize the method introduced in  [Balog et al., Phys. Rev. Lett. {\bf 129}, 210602 (2022)] to the $O(N)$ model. We study the large $N$ limit, as well as the cases $N=2$ and $N=3$ at the level of the Local Potential Approximation (LPA), and compare our results to Monte Carlo simulations.  We compute the entire family of universal scaling functions, obtained in the limit where the system size $L$ and the correlation length of the infinite system $\xi_\infty$ diverge, with the ratio $\zeta=L/\xi_\infty$ constant. We also generalize our results to the approach of criticality from the low-temperature phase where another infinite family of universal PDF exists.  We find that the LPA describes very well the functional form of the family of PDFs, once we correct for a global amplitude of the (logarithm of the) PDF and of $\zeta$.
\end{abstract}

\maketitle

\section{Introduction}

The emergence of universal behavior from strong correlations has intrigued physicists for decades. Near a continuous phase transition, the divergence of the correlation length effectively erases most microscopic details, leaving the system characterized by universal behaviors at large scales. These manifest as scaling laws for most observables. Crucially, broad classes of systems, categorized by factors such as the symmetry of the order parameter, dimensionality, and other global properties, exhibit identical universal properties. 

Furthermore, the diverging correlation length implies that the system's microscopic degrees of freedom are strongly correlated. This implies that close to criticality, the probability distribution function (PDF)  of the order parameter can be non-Gaussian. For a model involving a $N$-component microscopic field $\hbphi$ and described by a Hamiltonian ${\cal H}$, this PDF of a $d$-dimensional system of size $L$ is given by
\begin{equation}
\label{eq_def_P}
P(\hat \s=\s) =\mathcal N\int \DD\hbphi\, \delta\left(\s-\hat \s\right)\exp(-{\cal H}[\hbphi]/T),
\end{equation}
where $\hat \s=L^{-d}\int_\x \hbphi(\x)$ is the total spin per unit volume, $L$ the system size and $\mathcal N$ a normalization factor. We focus here on systems with short-range interactions in dimension $2<d<4$, for which the PDF is non-Gaussian at criticality. Noting that the choice of boundary conditions affects the shape of the PDF, we will consider here only periodic boundary conditions. Equivalently, one can study the rate function $I(\s)$, also known as the constraint effective potential in high-energy physics \cite{Fukuda1975,ORaifeartaigh1986,Mukaida1996}, defined as
\begin{equation}
    \label{def_I}
      I(\s) = -\lim_{L\to\infty} L^{-d} \ln P(\s).
\end{equation}
Standard scaling arguments imply that close to the critical temperature $T_c$, $I(\s)$, as a function of $\s$, $T-T_c$  and $L$ should obey a scaling law \cite{Bouchaud1990}
\begin{equation}
    I(\s,\xi_{\infty},L)=b^{-d} I(\s b^{(d-2+\eta)/2},\xi_{\infty}/b,L/b),
    \label{eq_scaleI}
\end{equation}
where $\eta$ the anomalous dimension, and we use the (infinite system) correlation length $\xi_{\infty}\propto (T-T_c)^{-\nu}$ to parametrize the distance to the critical point, $\nu$ being the correlation length critical exponent. This implies that close to criticality ($L\to\infty$, $\xi_\infty\to\infty$), $L^d I$ is a universal function of $\s L^{(d-2+\eta)/2}$ and $\zeta=L/\xi_\infty$. There is thus a whole family of critical PDFs parametrized by $\zeta$, corresponding to various ways of reaching the critical point and the thermodynamic limit. Furthermore, a standard argument by Privman and Fisher \cite{privman1984} implies that the amplitude of the scaling function of $L^d I$ is also universal.

The study of the shape of the PDF (or of the rate function) of such strongly correlated variables has a long history, especially using numerical methods and at $\zeta=0$, for the Ising model \cite{Binder1981,Bruce1992,Tsypin1994,Tsypin2000,Hilfer2003,Hilfer2005,Malakis2006,Xu2020} and for the $O(4)$ model in high energy physics \cite{Hasenfratz1987,Kuti1988,Hasenfratz1989,Gockeler1991,Gockeler1991a,Gockeler1991b,Dimitrovic1991,Gockeler1993}. This also includes the PDF of the 2D low-temperature XY model and its possible connection with turbulence \cite{BHP_1,BHP_2,Portelli_02,Archambault_1997,Joubaud2008}.
From a theoretical perspective, Wilson's renormalization group (RG) \cite{WilsonKogut_74} provides the correct framework to understand critical phenomena. By systematically coarse-graining the microscopic degrees of freedom, RG leads to an effective macroscopic description, where universality arises because microscopically different systems can exhibit the same large-scale behavior. This conceptual framework also finds a natural connection with probability theory, as RG can be interpreted as the flow of the probability distribution of block-spins \cite{Bleher1973,Bleher1975,Bleher1987,Bleher1988,Bleher1989,JonaLasino1975,Cassandro1978,JonaLasino2001}. In this sense, the RG provides a powerful tool for computing the properties of the order parameter near criticality, and in particular its PDF, in practice often relying on perturbative techniques or ad hoc methods \cite{Bruce1979,Rudnick1985,Eisenriegler1987,Hilfer1993,Hilfer1995,Esser1995,ChenDohm_96,Chen1998,Rudnick1998,Sahu2024}.

The functional RG (FRG) is a modern implementation of Wilson's idea that allows for non-perturbative approximation schemes, see \cite{Berges_02,Dupuis2021} for reviews. Recently, we have shown how to modify FRG to compute the rate function of the three-dimensional Ising model close to criticality. Here, we generalize the method to the $O(N)$ model and compare it to the exact calculation in large $N$ and to Monte Carlo simulations for $N=2$ and $N=3$. We show that the method is able to capture the shape of the whole family of rate functions,  once we correct for a global amplitude of the rate function and of $\zeta$.

The article is organized as follows. Section \ref{sec_Ninf} discusses the rate function in the large $N$ limit. In Sec.~\ref{sec_RG_rate}, we discuss the general formalism for computing the rate function within FRG, defining the scale-dependent constraint effective action (Sec.~\ref{sec_RG_eff_act}) and the LPA approximation (Sec. \ref{sec_RG_LPA}). We then describe how we solve numerically the LPA flow equations (Sec.~\ref{sec_comparisons_numeric_int}) and benchmark the numerical implementation on the large $N$ limit where LPA is exact (Sec.~\ref{sec_comparisons_benchm_Ninf}). Then in Sec.~\ref{sec_comparisons} we present the results of FRG at LPA for the rate functions $N=2$ and $3$ and compare them to Monte Carlo simulations. The conclusion is given in Sec.~\ref{sec_discussion}.

\section{Large $N$ limit}
\label{sec_Ninf}

In this section, we compute the rate function of the $O(N)$ model in the large $N$ limit. We start presenting the derivation of the effective potential (Gibbs free energy), which is standard, see for instance \cite{Moshe2003}. This is useful for two reasons. First, the calculation of the rate function is very similar to that of the effective potential. Second, it is necessary to have the expression for the effective potential to find the critical point and define the correlation length and thus control the parameter $L/\xi_\infty$, which parametrizes the rate functions. It will also allow us to discuss the differences between the rate function and the effective potential at finite size.

\subsection{Effective potential}
 \label{sec_eff_pot_Ninf}

The $O(N)$ model is defined by the Hamiltonian
\begin{equation}
    \mathcal H[\hbphi]=\int_\x \left(\frac{(\nabla\hbphi)^2}2+V\left(\hbphi^2/2\right)\right),
\end{equation}
where  $V$ is the microscopic potential, which is typically of the form $V(z) = r_0 z+\frac{u_0}{6 N}z^2$. The scaling in $N$ is chosen such that $V(z)$ is of order $N$ if  $z$ is of order $N$.

In dimension $d>2$, there is a second-order phase transition between a ferromagnetic ($\langle \hbphi\rangle\neq 0$) and a paramagnetic ($\langle \hbphi\rangle= 0$) phase when tuning the potential (e.g. $r_0$ at fixed $u_0$). Here, averages are performed using $\exp(-\mathcal H)$ as a Boltzmann weight.
The generating functional of correlation functions of the $O(N)$ model is defined as 
\begin{equation}
    \begin{split}
        \ZZ[\h]= \int \DD\hbphi\, e^{-\mathcal H[\hbphi]+\int_\x \h.\hbphi}.
    \end{split}
\end{equation}
In the large $N$ limit, it is convenient to introduce two auxiliary fields $\lambda(\x)$ and $\hrho(\x)$ such that 
\begin{equation}
    1=\int \DD \lambda\DD\hrho \,\exp\left\{-i\int_\x\lambda\left(\frac{\hbphi^2}2-\hrho\right)\right\}
\end{equation}
allowing us to reexpress the partition function of the $O(N)$ model as 
\begin{equation}
    \begin{split}
        \ZZ[\h]= \int \DD\hbphi\DD\lambda\DD\hrho\, e^{-\int_\x \left(\frac{(\nabla\hbphi)^2}2+i\lambda\frac{\hbphi^2}2\right)-\int_\x \left(V(\hrho)-i\lambda\hrho\right)+\int_\x \h.\hbphi}.
    \end{split}
\end{equation}
Writing the field $\hbphi=(\hsigma,\hbpi)$, with the $\hsigma$ direction along $\h$, and integrating out the $\hbpi$ fields, we finally obtain
\begin{equation}
        \ZZ[\h]= \int \DD\hsigma\DD\lambda\DD\hrho\, e^{-\HH_{\rm eff}[\hsigma,\lambda,\hrho]+\int_\x h \hsigma},
\end{equation}
with 
\begin{equation}\label{heff}
       \HH_{\rm eff}[\hsigma,\lambda,\hrho] = \int_\x \left(\frac{(\nabla\hsigma)^2}2+i\lambda\frac{\hsigma^2}2\right)+\int_\x \left(V(\hrho)-i\lambda\hrho\right)+\frac{N-1}2\Tr\log(g_\pi^{-1}),
\end{equation}
and with the correlation function $g_\pi$ of the $\hbpi$ fields satisfying $(-\nabla^2+i\lambda(\x))g_\pi(\x,\y)=\delta(\x-\y)$. 

We aim at computing the effective action $\Gamma[\bphi]$ (or Gibbs free energy), the Legendre transform of $\ln \ZZ[\h]$ with respect to $\h$, i.e.
\begin{equation}
    \Gamma[\bphi]= -\ln \ZZ[\h]+\int_\x \h.\bphi,
\end{equation}
with $\bphi(\x)=\langle \hbphi(\x)\rangle$ the magnetization in presence of the magnetic field $\h$.
Assuming that  $\hsigma\sim \sqrt{N}$, $\hrho\sim N, \lambda\sim1$ and $V(\hrho)\sim N$ for $N$ large, all terms in Eq.~\eqref{heff} are of order $N$ and the functional integral can be evaluated at the saddle-point as $N\to\infty$
\begin{equation}
\begin{split}
\frac{\delta \HH_{\rm eff}}{\delta \lambda(\x)}\bigg|_{\lambda=\lambda_0}=0, \\
\frac{\delta \HH_{\rm eff}}{\delta \hrho(\x)}\bigg|_{\hrho=\hrho_0}=0.
\end{split}
\end{equation}
Furthermore, the effective action reads in this approximation
\begin{equation}
    \Gamma[\bphi]= \HH_{\rm eff}[\bphi, \lambda_0,\hrho_0],
\end{equation}
where we have used the $O(N)$ invariance to replace $\hsigma$ by $\bphi$ when performing the Legendre transform.

In uniform magnetization $\bphi(\x)=\bphi$, $\Gamma[\bphi]=L^d U(\rho)$ with $\rho=\bphi^2/2$, and the saddle-point solutions $\lambda_0$ and $\hrho_0$ are also uniform (and dependent on $\rho$). One shows that $i\lambda_0=U'(\rho)$, primes meaning derivatives with respect to $\rho$, while the saddle-point equations read
\begin{equation}
\begin{split}
i\lambda_0&=V'(\hrho_0),\\
\rho &= \hrho_0-\frac{N}2\frac{1}{L^d}\sum_\q\frac{1}{\q^2+W(\rho)},
\end{split}
\label{eq_gap_bare}    
\end{equation}
with $W\equiv U'$, and the sum over momenta is understood as being regularized in the ultra-violet (UV), that is, for large momenta $\vert\q\vert$.

Finally, it is convenient to introduce the inverse of the function $V'$, which we call $f_\Lambda(W)$ for later convenience (see Sec.~\ref{sec_comparisons_benchm_Ninf}), i.e. $V'(f_\Lambda(W))=W$.
The gap equation can be rewritten as an implicit equation for $\rho$ seen as a function of $W$,
\begin{equation}
\label{eq_gap_bare1}
\rho = f_\Lambda(W)-\frac{N}2\frac{1}{L^d}\sum_\q\frac{1}{\q^2+W}.
\end{equation}

In the following, we regularize the sum over momenta by introducing an exponential factor $e^{-\q^2/\Lambda^2}$, which allows for using Schwinger's representation of the propagator and perform explicitly the sum over momenta \cite{Moshe2003}
\begin{equation}
\begin{split}
\frac{1}{L^d}\sum_\q \frac{1}{\q^2+W} \ \to\ \frac{1}{L^d}\sum_\q \frac{e^{-\q^2/\Lambda^2}}{\q^2+W}&= \frac{1}{L^d}e^{W/\Lambda^2}\int_{\Lambda^{-2}}^\infty ds \sum_\q e^{-s(\q^2+W)},\\
&= \frac{1}{L^d}\int_{\Lambda^{-2}}^\infty ds\, e^{-s W} \uvtheta^d(s 4\pi /L^2)+\mathcal O(W/\Lambda^2),\\
&= \frac{1}{L^{d-2}}\int_{\frac{4\pi}{\Lambda ^2 L^2}}^\infty \frac{ds}{4\pi}e^{-s \frac{L^2W}{4\pi}}\uvtheta^d(s)+\mathcal O(W/\Lambda^2),
\end{split}
\end{equation}
 where we have neglected non-universal terms of order $W/\Lambda^2$ (which can be absorbed into $f_\Lambda$) and $\uvtheta(s)=\sum_{n\in \mathbb Z}e^{-s \pi n^2}$ is a Jacobi theta function, which obeys in particular $\uvtheta(s)=s^{-1/2}\uvtheta(1/s)$. The UV divergence of the momentum integral translates into the divergence $s^{-d/2}$ at small $s$ for $2<d<4$. Adding and subtracting such a term and taking the limit $\Lambda\to\infty$ in the integral, we obtain
\begin{equation}
\begin{split}
\frac{1}{L^d}\sum_\q \frac{e^{-\q^2/\Lambda^2}}{\q^2+W}&=\frac{2}{d-2}\frac{\Lambda^{d-2}}{(4\pi)^{d/2}}+\frac{1}{L^{d-2}}\int_{0}^\infty \frac{du}{4\pi}\left(e^{-u \frac{L^2W}{4\pi}}\uvtheta^d(u)-u^{-d/2}\right)+\mathcal O(W^2/\Lambda^2).
\end{split}
\end{equation}
It will be convenient to include the UV-dependent constant into the function $f_\Lambda$, defining $\hat f_\Lambda(W)= f_\Lambda(W)-\frac{N}{2(d-2)}\frac{\Lambda^{d-2}}{(4\pi)^{d/2}}$.

The function 
\begin{equation}
\label{eq_F_definition}
    F_d(z)=-\frac12\int_{0}^\infty \frac{du}{4\pi}\left(e^{-u z}\uvtheta^d(u)-u^{-d/2}\right)
\end{equation} 
is shown in Fig. \ref{fig:F_vs_hF}. It is monotonically increasing and interpolates between $-1/(8\pi z)$ at small $z$ (corresponding to the term $\q=0$ of the sum over momenta in Eq.~\eqref{eq_gap_bare}) and $-\Gamma(1-d/2)\frac{z^{\frac{d-2}{2}}}{8\pi}+\mathcal{O}(e^{-2\sqrt{\pi z}})$ at large $z$ (corresponding to the thermodynamic limit of the sum up to exponentially small correction terms).

The critical point is found by taking the thermodynamic limit, $L\to\infty$, for which the effective potential becomes independent of $L$, $W\to W_{\infty}$, and the gap equation becomes in this limit
\begin{equation}
\label{eq_gap_TL}
\rho = \hat f_\Lambda(W_{\infty})+N c_d W_{\infty}^{\frac{d-2}2},
\end{equation}
with
\begin{equation}
    c_d=-\frac{\Gamma(1-d/2)}{2(4\pi)^{d/2}}>0.
\end{equation}
Note that $\hat f_\Lambda(W)$ is a regular function of $W$ at small $W$ (the regime of interest, since the correlation length $\xi$ is proportional to $W^{-1/2}$), and the term proportional to $W_\infty^{(d-2)/2}$ is therefore dominant compared to a term of order $W_\infty$. We can therefore expand $\hat f_\Lambda(W_\infty)=-N\delta+\mathcal O(W_\infty)$ and the gap equation in the thermodynamic limit becomes
\begin{equation}
\label{gap_td_limt}
\begin{split}
\rho/N = -\delta+c_d W_{\infty}^{\frac{d-2}2}.
\end{split}
\end{equation}
Note that $\delta$ depends on the Hamiltonian and is thus non-universal (e.g. $\delta=3r_0/u_0+\Lambda^{d-2}/((4\pi)^{d/2}(d-2))$ for a $(\hat\bphi^2)^2$ theory with the regularization described above).

The critical point is found by solving $W_{\infty}(0)=0$ (minimum of the free energy at zero magnetization and infinite correlation length $\xi_\infty^2=1/W_\infty(0)$) and thus corresponds to $\delta=0$. 
The system is disordered for $\delta>0$ (minimum at zero magnetization and finite correlation length), with correlation length diverging as $\xi_{\infty}(\delta)= (\delta/c_d)^{-\nu}$ with $\nu=1/(d-2)$ as expected. For $\delta<0$, the minimum of the free energy is at $\rho_0=N\delta$, from which we obtain the critical exponent $\beta=1/2$. Note that the effective potential is not defined for $\rho<\rho_0$ since by construction $|\langle \hbphi\rangle_{h,\infty}|^2\geq 2\rho_0$ for all possible magnetic fields, and the relationship between $h$ and $\phi$ cannot be inverted (when doing the Legendre transform) for smaller magnetization.

For finite-size systems, in the limit of small $W$ (i.e. $\hat f_\Lambda(W)=-N\delta+\mathcal O(W)$), the gap equation reads
\begin{equation}
\begin{split}
\rho = -N\delta +\frac{N}{L^{d-2}}F_d\left(\frac{L^2 W}{4\pi}\right).
\end{split}
\end{equation}
The system starts showing critical-like behavior for $\delta$ small enough and large enough size $L$. As noted in \cite{Balog22}, there are many inequivalent ways to take the limit $\delta\to 0$ and $L\to \infty$, leading to different scaling functions. We find it convenient to parametrize these in terms of $\zeta={\rm sign}(\delta) L/\xi_\infty(|\delta|)$, which is negative (positive) when criticality is obtained coming from the ordered (disordered) phase. Note that while the relationship between $\delta$ and $\xi_\infty$ depends on a non-universal amplitude, $\zeta$ is free of such ambiguity, which makes it a good parametrization.

At finite size $L$, $W$ cannot vanish for any $\rho$ (since we would have a divergence at $\q=0$ in the sum in Eq.~\eqref{eq_gap_bare}), and thus $W(\rho)>0$ and the effective potential is strictly convex. Its minimum is thus at $\rho=0$ for all $\delta$, corresponding to the fact that the magnetization of a finite system vanishes, since spontaneous symmetry breaking can only happen in infinite systems. The gap equation reads
\begin{equation}
\begin{split}
F_d\left(\frac{L^2 W}{4\pi}\right) =\frac{L^{d-2}\rho}N+\Delta,
\end{split}
\end{equation}
where we have introduced $\Delta=L^{d-2} \delta$, related to $\zeta$ by
\begin{equation}
    \Delta= {\rm sign} (\delta)c_d|\zeta|^{d-2}.
    \label{eq_delta_Ninf}
\end{equation}
We find that $L^2W$ is a universal function of $\frac{L^{d-2}\rho}N$, and $\Delta$ plays the role of a shift in $\rho$.

Deep in the disordered phase for finite but large $L$ (i.e. $\delta>0$ and $\delta\propto O(1)$ independent of $L$, and $L$ large), $W$ differs from $W_\infty$ by an exponentially small correction, see Eq.~\eqref{eq_F_definition} and below. 
Deep in the ordered side ($\delta<0$ and $\vert\delta\vert\propto O(1)$ independent of $L$, and $L$ large), we find that $W(0)\propto L^{-d}$. Furthermore, we find that for $\rho<\rho_0$, $W(\rho)\simeq \frac{N}{2L^d(\rho_0-\rho)}$.  
Finally, for $|\rho_0-\rho| \propto L^{-d+2}$, $W(\rho)\propto L^{-2}$.
The shape of the effective potential in the ordered phase can thus be expressed as $U(\rho)=\frac{N}{L^d}G\left(L^{d-2}(\rho-\rho_0)/N\right)$ with 
\begin{equation}
G(x) \propto
\begin{cases}
\log|x| & \text{if  $x < 0$ and } |x|\gg 1,\\
x & \text{if } |x|\ll 1, \\
x^{\frac{d}{d-2}} & \text{if $x > 0$ and } x \gg 1. \\
\end{cases}    
\end{equation}

For all $\Delta$, $W(\rho)\propto \rho^{\frac{2}{d-2}}$ at large field, corresponding to $U(\rho)\propto \rho^{\frac{d}{d-2+\eta}}$ with $\eta=0$ for large $N$.\footnote{This is valid as long as $\rho$ is not too large, i.e. as long as the universal part dominates compared to $\hat f_\Lambda$. At very large field, we recover $U(\rho)=V(\rho)$, the microscopic (and thus non-universal) potential.}

\subsection{Rate function}
\label{sec_rate_Ninf}

The derivation for the rate function is similar to that for the effective potential explained in Sec. \ref{sec_eff_pot_Ninf}, see also \cite{Mukaida1996}. Starting from Eq.~\eqref{eq_def_P} and using the exponential representation of the delta-function, $\delta(\mathbf{z})\propto \int d\h\, e^{i \h.\mathbf{z}}$,\footnote{See \cite{Balog2024} for an alternative calculation using a different exponentiation of the delta-function.} such that 
\begin{equation}
P(\hat \s= \s) =\mathcal N'\int  d\h\, e^{-i L^d \h.\s}\int \DD\hbphi\, e^{-{\cal H}[\hbphi]+i \int_\x\h.\hbphi}
\label{eq_P_int}
\end{equation}
with $\mathcal N'$ a new normalization factor. Introducing two auxiliary fields $\clambda(\x)$ and $\hrho(\x)$, Eq.~\eqref{eq_P_int} reduces to
\begin{equation}
    \begin{split}
        P(\s)= \int \DD\hbphi\DD\clambda\DD\hrho d\h\, e^{-\int_\x \left(\frac{(\nabla\hbphi)^2}2+i\clambda\frac{\hbphi^2}2\right)-\int_\x \left(V(\hrho)-i\clambda\hrho\right)+i\h.\int_\x(\s-\hbphi)}.
    \end{split}
\end{equation}
We stress that here, $\clambda$, $\hrho$ and $\bphi$ depend on $\x$, whereas $\h$ and $\s$ do not (in particular, the integral over $\h$ is not a functional integral). Calling $\csigma$ the direction of $\hbphi$ along $\s$, and splitting $\h$ into $h_\sigma$ and $\h_\pi$, the integral over the $\hbpi$ fields followed by that over $\h_\pi$ gives
\begin{equation}
    \begin{split}
        P(\s)&= \int \DD\csigma\DD\clambda\DD\hrho d\h\, e^{-\HH_{eff}[\csigma,\clambda,\hrho]+i h_\sigma\int_\x(s-\csigma)-\frac{N-1}2\Tr\log(\check{g}_\pi^{-1})-\h_\pi^2\int_{\x,\y}\check{g}_\pi(\x,\y)},\\
        &=\int \DD\csigma\DD\clambda\DD\hrho dh_\sigma \, e^{-\HH_{eff}[\csigma,\clambda,\hrho]+i h_\sigma\int_\x(s-\csigma)-\frac{N-1}2\Tr\log(\check{g}_\pi^{-1})-\frac{N-1}2\log\left(\int_{\x,\y}\check{g}_\pi(\x,\y)\right)},
    \end{split}
\end{equation}
where  $(-\nabla^2+i\clambda(\x))\check{g}_\pi(\x,\y)=\delta(\x-\y)$.
The saddle-point equations read, assuming that the solution is in constant fields,
\begin{equation}
    \begin{split}
        \csigma&=s,\\
        i\clambda \csigma &= i h_\sigma,\\
        i\clambda &= V'(\hrho),\\
        \frac{\csigma^2}2 &=\hrho-\frac{N}{2L^d}\sum_{\q\neq 0}\frac{1}{\q^2+i\clambda}.
    \end{split}
\end{equation}
In the last equation, the subtraction of the $\q=0$ term comes from the derivative with respect to $i\clambda$ of $\log(\int_{\x,\y}g_\pi(\x,\y))=\log(L^d/i\clambda)$ in constant field.

Writing $\log(P(\s))=-L^d I(\crho)$ with $\crho=\s^2/2$, one shows that at the saddle-point, $i\clambda =\cW(\crho)\equiv I'(\crho)$, and thus the gap equation of the rate function is
\begin{equation}
\label{eq_gap_rate_proto}
    \crho=f_\Lambda(\cW)-\frac{N}{2L^d}\sum_{\q\neq 0}\frac{1}{\q^2+\cW},
\end{equation}
where $f_\Lambda$ is the same function as in Eq.~\eqref{eq_gap_bare1}. Equation \eqref{eq_gap_rate_proto} is very similar to the gap equation \eqref{eq_gap_bare1}, the only difference being the absence of the $\q=0$ term in the sum over momenta.
 We note right away that in the limit $L\to\infty$, the absence of this term does not matter, and the gap equations of $W$ and $\cW$ become identical, hence we recover the standard result that in the thermodynamic limit $I(\crho)=U(\rho=\crho)$.

Regularizing the gap equation as above, absorbing terms of order $\Lambda^{d-2}$ into $f_\Lambda$ which we rename $\hat f_\Lambda$, and focusing on the critical regime where we can approximate $\hat f_\Lambda(\cW)\simeq -N\delta$, one can express the gap equation as 
\begin{equation}
\label{eq_gap_rate}
\begin{split}
\crho = -N\delta +\frac{N}{L^{d-2}}\cF_d\left(\frac{L^2 \cW}{4\pi}\right),
\end{split}
\end{equation}
where 
\begin{equation}
\begin{split}
\cF_d(z) = -\frac12\int_{0}^\infty \frac{du}{4\pi}\left(e^{-u z}(\uvtheta^d(u)-1)-u^{-d/2}\right).
\end{split}
\label{eq_def_cF}
\end{equation}
The only difference with Eq.~\eqref{eq_F_definition}
is the presence of the $-1$ in the integrand, coming from subtracting the $\q=0$ term.

\begin{figure}
    \centering
    \includegraphics[width=8cm]{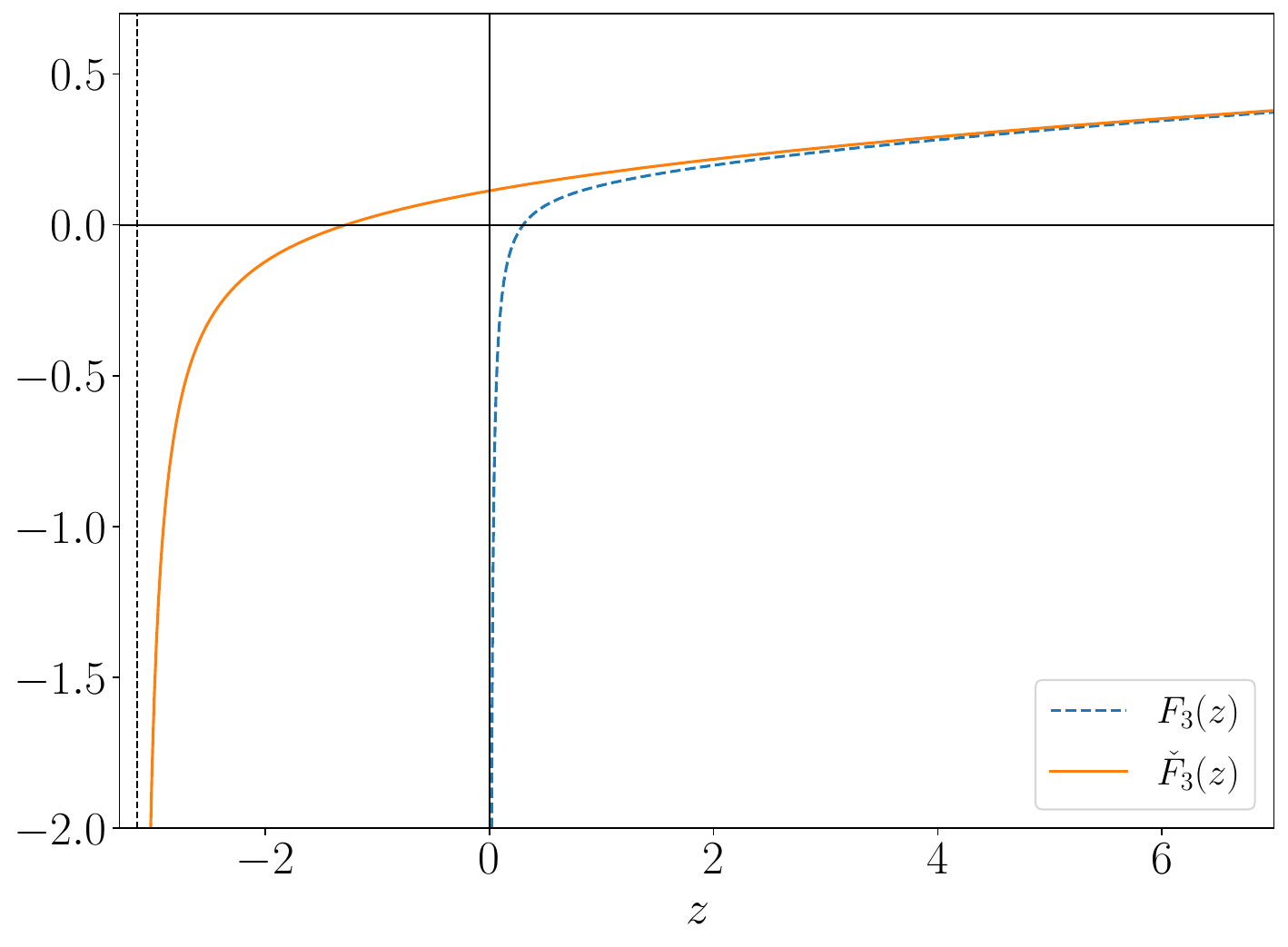}
    \caption{
    Comparison between the universal functions $F_d$, Eq.~\eqref{eq_F_definition},  and $\cF_d$, Eq.~\eqref{eq_def_cF}, for $d=3$. The vertical dashed line shows the position of the pole of  $\cF_d(z)$ at  $z=-\pi$. The pole of $F_d(z)$ is at $z=0$. 
    }
    \label{fig:F_vs_hF}
\end{figure}

That the thermodynamic limit is the same for both $W$ and $\cW$ translates into $\cF_d(z) \to F_d(z)$ for $z\gg 1$. On the other hand, since the term $\q=0$ is absent from the sum over momenta, $\cF_d$ does not have a pole at $z=0$, but instead has a pole at a {\it negative} value of $z=-\pi$, $\cF_d(z)\simeq -\frac{2d}{8\pi}\frac{1}{\pi+z}$. This allows for a non-convex rate function at finite size. Finally, let us note that $\cF_d$ is a monotone function and that $\cF_d(0)>0$ for all $2<d<4$, which implies that $\cF_d(\check z_0)=0$ for some $\check z_0\in ]-\pi,0[$. Fig.~\ref{fig:F_vs_hF} shows the comparison between  $\cF_d$ and  $F_d$ for $d=3$.

\begin{figure}
    \centering
    \includegraphics[width=8cm]{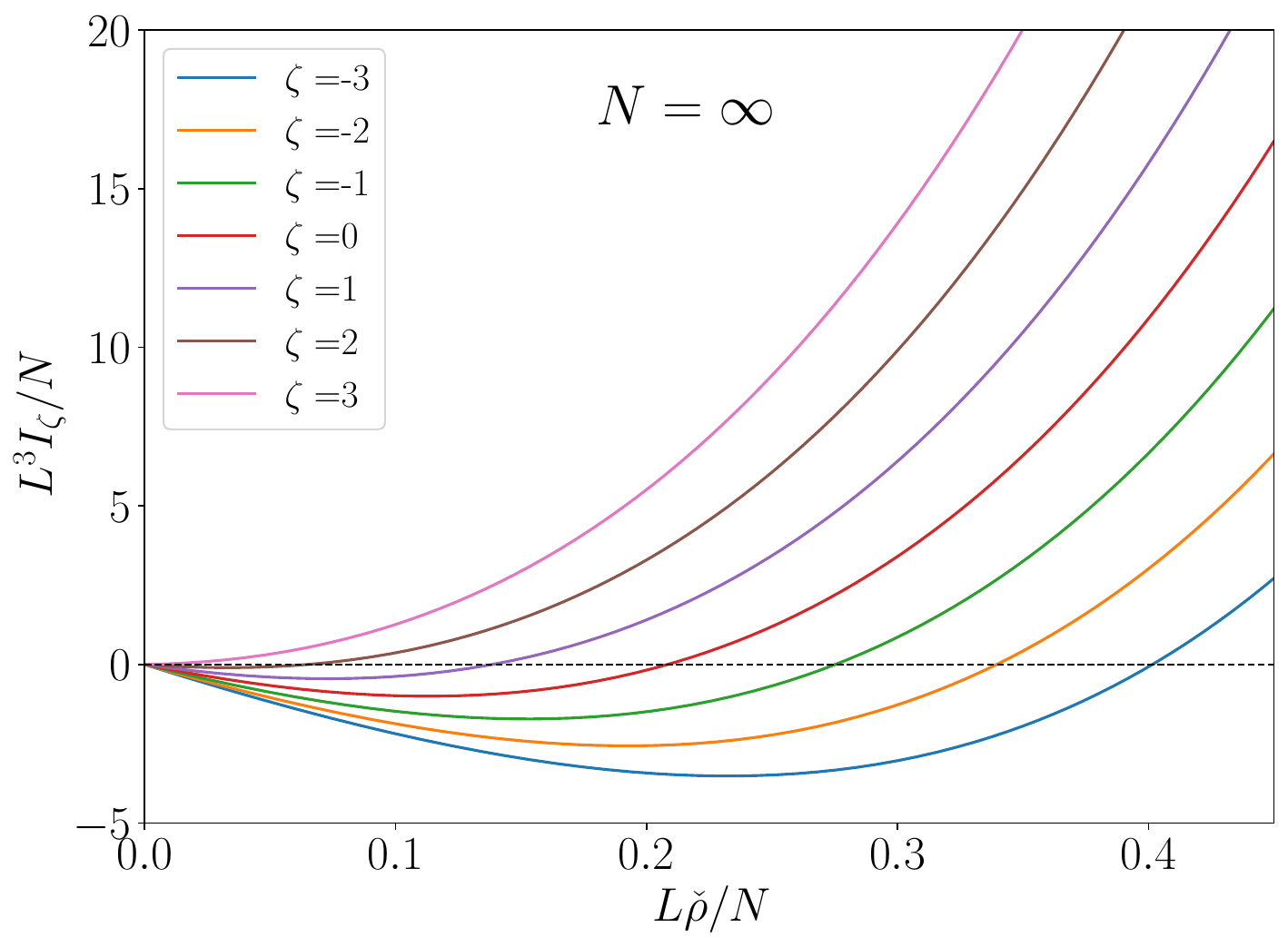}
    \caption{
    Rate function in the large $N$ limit for $d=3$ for several values of $\zeta$.  The rate function changes convexity at $\zeta=\zeta_{c,\infty}\approx 2.84$.}
    \label{fig:I_crit_Ninf}
\end{figure}

In the critical regime, at fixed $\Delta$, we find
\begin{equation}
\begin{split}
\cF_d\left(\frac{L^2 \cW}{4\pi}\right)=\frac{L^{d-2}\crho}{N}+\Delta,
\end{split}
\label{eq_hatF_definition}
\end{equation}
and the rate function is non-convex as long as $\cF_d(z)=\Delta$ has a solution for negative $z$, i.e. up to  $\Delta_c=\cF_d(0)>0$ (which, using Eq.~\eqref{eq_delta_Ninf}, translates into a critical $\zeta$, $\zeta_{c,\infty}\simeq2.8373\ldots$). Indeed, we find that $\cW$ vanishes at a finite value of $\crho=\crho_{0}=\frac{N }{L^{d-2}}\left(\cF_d(0)-\Delta\right)$, and $\cW(\crho)<0$ for $\crho\in[0,\crho_{0}[$ and positive otherwise. Furthermore, we directly find that the rate function is given by a universal function of $\frac{L^{d-2}\crho}{N}$,
\begin{equation}
    I(\crho,\delta)=N L^{-d}I_\zeta\left(\frac{L^{d-2}\crho}{N}\right),
\end{equation}
with $I_\zeta(x)$ such that
\begin{equation}
    \cF_d\left(\frac{L^2I'_\zeta(x)}{4\pi}\right)=x+\Delta,
\end{equation}
with $\Delta$ related to $\zeta$ by Eq.~\eqref{eq_delta_Ninf}. Fig.~\ref{fig:I_crit_Ninf} shows the rate functions in the critical regime for several $\zeta$.

Deep in the disordered phase, $\delta>0$ and $\xi_\infty\ll L$, we recover $\cW(\crho)\simeq W_\infty(\rho=\crho)$, a convex function.
Deep in the ordered phase, $\delta<0$ and $L$ large, with a finite magnetization $\frac{\rho_0}{N}=- \delta>0$ and $\frac{\rho_0}{N}\propto O(1)$ in the thermodynamic limit, the minimum of the rate function is at $\frac{\crho}{N}=\frac{\rho_0}{N}+\frac{1}{L^{d-2}}\cF_d(0)$ (which of course tends to $\rho_0$ for $L\to \infty$), and it is non-convex for any finite $L$. For small enough $\crho$, we find
\begin{equation}
    \cW(\crho)\simeq -\frac{4\pi^2}{L^2}+\frac{2d}{L^d(\rho_0-\crho)}.
    \label{eq_largeN_order}
\end{equation}
Therefore, the typical size of the free-energy-density barrier between $\crho=0$ and $\crho=\rho_0$ is $\rho_0/L^2$ corresponding to an energy of order $L^{d-2}\rho_0$. This corresponds to the energy of a spin-wave excitation of wavelength of order $L$, needed to lower the magnetization below the equilibrium one.~\footnote{Interestingly, a result similar to Eq.~\eqref{eq_largeN_order} has been obtained in \cite{pelaez2016} when studying the ordered phase in the large $N$ limit with FRG. There, the role of $L$ is played by the RG scale $k^{-1}$.}

\section{Renormalization group approach to the rate function calculation}
\label{sec_RG_rate}

There is no known method to solve the $O(N)$ model at finite $N$. In this section, we generalize the FRG approach devised in \cite{Balog22} to compute PDFs for Ising to the case of the $O(N)$ model.

\subsection{Definition of the effective actions $\Gamma_{M,k}$}
\label{sec_RG_eff_act}

The derivation of the FRG for the calculation of rate functions follows closely the original derivation \cite{Wetterich1993,Wetterich1993a,Wetterich1993b}, see  \cite{Berges_02,Dupuis2021} for reviews of the method. In a nutshell, given the Hamiltonian of a theory $\mathcal{H}[\hbphi]$, a family of models is defined by introducing an addition to the Hamiltonian, $\Delta\mathcal{H}_k=\frac{1}{2}\hbphi.R_k.\hbphi$ with $R_k(\x,\y)$ an infrared regulator, parametrized by a momentum RG scale $k$. It is chosen such that the integration over modes with momenta $|q|<k$ is suppressed while it leaves the modes $|q|>k$ unchanged. Furthermore, we impose that $R_{k=0}=0$ while $\lim_{k\to \Lambda}R_k$ is such that the modified model is exactly solvable in this limit.
Defining a family of scale-dependent partition functions $\mathcal{Z}_k[\h]$ and a scale-dependent free energy functional $-\ln(\mathcal{Z}_k)$, its (modified) Legendre transform $\Gamma_k[\bphi]$ is the object of interest. This scale-dependent effective action $\Gamma_k[\bphi]$ obeys an exact RG flow equation, which can be approximated non-perturbatively.

To compute the rate function, one can follow a similar route, where now the (scale-dependent) partition function is modified to include the delta function found in Eq.~\eqref{eq_def_P},
\begin{equation}
\label{eq_ZMk_definition}
\mathcal Z_{M,k}[\h] =\int \DD\hbphi\, e^{-\mathcal H_M[\hbphi] + \h.\hbphi - \frac12 \hbphi.R_k.\hbphi}
\end{equation}
where ${\cal H}_M[\hbphi]={\cal H}(\hbphi)+\frac {M^2}2 [\int_\x (\hbphi(\x)-\s)]^2$. In the limit $M\to\infty$, the second term in ${\cal H}_M$ becomes a delta function, and we recover the constraint $L^{-d}\int_\x \hbphi(\x)=\s$.
This implementation of the constraint is different from that of Sec.~\ref{sec_rate_Ninf}, but it is better suited for the present purpose.

Introducing $\bphi(\x)=\frac{\delta  \ln \mathcal Z_{M,k}}{\delta \h(\x)}$, we define the (modified) Legendre transform of $\ln \mathcal Z_{M,k}$ by
\begin{equation}
\Gamma_{M,k}[\bphi]=-\ln \mathcal Z_{M,k}[\h]+\h.\bphi-\frac12\bphi. R_k.\bphi-\frac {M^2}2 \left[\int_\x (\bphi(\x)-\s)\right]^2.
\label{eq_def_GammaA_LT}
\end{equation}
Using 
\begin{equation}
\frac{\delta\Gamma_{M,k}}{\delta\bphi(\x)}=\h(\x)-\int_\y R_k(\x,\y)\bphi(\y)-M^2\int_\y(\bphi(\y)-\s),
\end{equation}
allows us to write
\begin{equation}
\begin{split}
e^{-\Gamma_{M,k}[\bphi]}=\int \DD\hat\phi\, e^{-{\cal H}(\hbphi)+\frac{\delta\Gamma_{M,k}}{\delta\bphi}.(\hbphi-\bphi) - \frac12 (\hbphi-\bphi).R_{k}.(\hbphi-\bphi)} e^{-\frac {M^2}2 [\int_\x (\hbphi(\x)-\bphi(\x))]^2}.
\end{split}
\label{eq_GammaMk}
\end{equation}
 Note that although it is not explicit from its definition, Eq.~\eqref{eq_def_GammaA_LT},  $\Gamma_{M,k}[\bphi]$ is independent of $s$ as can be checked from Eq.~\eqref{eq_def_GammaA_LT_nice} or by deriving Eq.~\eqref{eq_def_GammaA_LT} with respect to $\s$.

When evaluated in a constant field $\bphi(\x)=\s$ and at $k=0$,
\begin{equation}
\frac{\delta\Gamma_{M,k=0}}{\delta\bphi(\x)}\bigg|_{\bphi(\x)=\s}=\h,
\end{equation}
 is a constant, by translation invariance, such that $\langle\hbphi(\x)\rangle =\s$. Thus, 
\begin{equation}
e^{-\Gamma_{M,k=0}[\bphi(\x)=\s]}=\int \DD\hbphi\, e^{-{\cal H}(\hbphi) -\frac {M^2}2[\int_\x(\hbphi-\s)]^2 +\h.\int_\x(\hbphi-\s)},
\end{equation}
and therefore
\begin{equation}
\begin{split}
    \lim_{M\to\infty} e^{-\Gamma_{M,k=0}[\bphi(\x)=\s]}\propto&\int \DD\hat\phi\, \delta\left(\s-L^{-d}\int_\x \hbphi\right)e^{-{\cal H}(\hbphi)},\\
 \propto&\ P(\s).
\end{split}
\end{equation}
This directly proves that $\lim_{M\to\infty}\Gamma_{M,k=0}[\bphi(\x)=\s]=L^d I(\s)$. 
At finite $k$, it is convenient to define $\check \Gamma_k[\bphi]=\lim_{M\to\infty}\Gamma_{M,k}[\bphi]$, and we interpret $L^{-d}\check \Gamma_k[\bphi(\x)=\s]=I_k(\s)$ as a scale-dependent rate function.

Defining $R_{M,k}(\x,\y)=R_{k}(\x,\y)+M^2$, Eq.~\eqref{eq_GammaMk} can be rewritten as
\begin{equation}
e^{-\Gamma_{M,k}[\bphi]}=\int \DD\hat\phi\, e^{-{\cal H}(\hbphi)+\frac{\delta\Gamma_{M,k}}{\delta\bphi}.(\hbphi-\bphi) - \frac12 (\hbphi-\bphi).R_{M,k}.(\hbphi-\bphi) },
\label{eq_def_GammaA_LT_nice}
\end{equation}
and therefore $\Gamma_{M,k}[\bphi]$ can be interpreted as the scale-dependent Gibbs free energy of a system with  Hamiltonian $\cal H$ regularized by $R_{M,k}$. Thus, the exact flow equation of $\Gamma_{M,k}$ can be found similarly to that of the standard scale-dependent effective action $\Gamma_k=\Gamma_{M=0,k}$ \cite{Balog22}
\begin{equation}
\partial_k \Gamma_{M,k}[\phi]=\frac{1}{2}\Tr\int_{\x,\y}\partial_k R_{M,k}(\x,\y) \left(\Gamma_{M,k}^{(2)}+R_{M,k}\right)^{-1}(\x,\y),
\end{equation} 
where the trace is over the $O(N)$ indices, and the elements of the matrix $\Gamma_{M,k}^{(2)}$ are defined as
\begin{equation}
\left(\Gamma_{M,k}^{(2)}\right)_{ij}=\Gamma_{ij,M,k}^{(2)}[\x,\y;\bphi]=\frac{\delta^2\Gamma_{M,k} }{\delta\phi_i(\x)\delta\phi_j(\y))}.
\end{equation}

The initial condition of the flow corresponds to $k=\Lambda$, for which $\lim_{k\to\Lambda}R_k(\q)\to\infty$. A saddle-point analysis gives
\begin{equation}
\lim_{k\to\Lambda}\mathcal Z_{M,k}[\h] \simeq e^{-\mathcal H_M[\bphi] + \h.\bphi - \frac12 \bphi.R_\Lambda.\bphi},
\end{equation} 
 with $\bphi[\h]$ such that 
\begin{equation}
\frac{\delta\mathcal H_M}{\delta\bphi(\x)} - \h(\x) + \int_\y R_\Lambda(\x,\y)\bphi(\y)=0.
\end{equation} 
Performing the Legendre transform and using Eq.~\eqref{eq_def_GammaA_LT}, we obtain
\begin{equation}
\lim_{k\to\Lambda}\Gamma_{M,k}[\bphi]=\mathcal H[\bphi].
\end{equation} 

Finally, note that the effective action $\Gamma[\bphi]$ of the original model, defined as the (true) Legendre transform of $\ln \mathcal Z_{M=0,k=0}[\h]$, is recovered from its scale-dependent counterpart $\Gamma_{k}[\bphi]=\Gamma_{M=0,k}[\bphi]$ by taking the limit $k\to0$. On the contrary, when $M\to\infty$, $\check \Gamma[\bphi]=\check\Gamma_{k=0}[\bphi]$ is not a true Legendre transform even in this limit, due to the $M$ term in Eq.~\eqref{eq_def_GammaA_LT}, and can therefore be non-convex. 

\subsection{Local Potential Approximation}
\label{sec_RG_LPA}

The Local Potential Approximation (LPA) corresponds to approximating the modified scale-dependent effective average action by the ansatz
\begin{equation}
\Gamma_{M,k}[\bphi]=\int_\x\left(\frac{(\nabla\bphi)^2}{2}+U_{M,k}(\rho)\right),
\end{equation} 
with $\rho(\x)=\bphi(\x)^2/2$, analogous to the LPA for the standard scale-dependent effective action \cite{Dupuis2021}. Note that $I_k = \lim_{M\to\infty}U_{M,k}$ and that $U_k = \lim_{M\to 0}U_{M,k}$.

Within this ansatz, the inverse propagator in momentum space and in constant field $\bphi$ reads
\begin{equation}
\Gamma_{ij,M,k}^{(2)}(\q;\bphi)=\delta_{ij}(\q^2+U_{M,k}'(\rho))+\phi_i\phi_j U_{M,k}''(\rho),
\end{equation} 
allowing to write explicitly the flow equation for $\Gamma_{M,k}$.
The propagator reads
\begin{equation}
    (\Gamma_{M,k}^{(2)}(\q;\bphi)+R_k(\q))^{-1}_{ij} = \frac{\phi_i\phi_j}{2\rho} \frac{1}{\q^2+U_{M,k}'(\rho)+2\rho U_{M,k}''(\rho)}+\left(\delta_{ij}-\frac{\phi_i\phi_j}{2\rho}\right)\frac{1}{\q^2+U_{M,k}'(\rho)}.
\end{equation}
In this approximation, the only flowing quantity is  $U_{M,k}$, and its flow equation reads
\begin{equation}
\label{flow_u_lpa}
\begin{split}
\partial_k U_{M,k}=\frac{1}{2L^d}\sum_\q\partial_k R_{k}(\q)\left(
%&
\frac{N-1}{\q^2+R_k(\q)+M^2\delta_{\q,0}+U'_{M,k}}
%\right.\\ \left.&
+\frac{1}{\q^2+R_k(\q)+M^2\delta_{\q,0}+U'_{M,k}+2\rho U''_{M,k}}\right),
\end{split}
\end{equation} 
where the trace over the $O(N)$ indices has been performed.

In the limit $M=0$, we recover the flow equation of the effective potential $U_k(\rho)$ (at finite size with periodic boundary conditions \cite{Fister2015})
\begin{equation}
\begin{split}
\partial_k U_{k}(\rho)=&\frac{1}{2L^d}\sum_\q\partial_k R_{k}(\q)\Bigg(\frac{N-1}{\q^2+R_k(\q)+U'_{k}}+\frac{1}{\q^2+R_k(\q)+U'_{k}+2\rho U''_{k}}\Bigg),
\end{split}
\end{equation} 
while in the limit $M\to\infty$ the zero-momentum contribution is suppressed, and the equation for $I_k$ reads (recall that $\rho$ is half of the average magnetization squared, while the argument of the rate function is $\crho=\s^2/2$)
\begin{equation}
\label{flow_i_lpa}
\begin{split}
\partial_k I_{k}(\crho)=&\frac{1}{2L^d}\sum_{\q\neq 0}\partial_k R_{k}(\q)\Bigg(\frac{N-1}{\q^2+R_k(\q)+I'_k}+\frac{1}{\q^2+R_k(\q)+I'_{k}+2\crho I''_{k}}\Bigg).
\end{split}
\end{equation} 
In the thermodynamic limit $L\to\infty$, the two sums converge to the same integral, and we recover the equivalence between the two, $I_k(\crho)=U_k(\rho=\crho)$.

To study the critical behavior, it is convenient to introduce the dimensionless momenta $\tilde \q=\q/k$ and fields
%\footnote{Here we use the standard FRG notation using a tilde for dimensionless fields in units of the RG scale $k$. Everywhere else, we use this notation to define dimensionless fields in units of the system size $L$. It should be clear from the context how this notation should be interpreted.}
$\tilde \rho=k^{-(d-2+\eta)}\rho$ and $\tcrhos=k^{-(d-2+\eta)}\crho$, and  $\tilde U_{k}(\tilde\rho)=k^{-d}U_{k}(\tilde\rho k^{d-2+\eta})$ and $\tilde I_{k}(\tcrhos )=k^{-d}I_{k}(\tcrhos k^{d-2+\eta})$. Here $\eta$ is the anomalous dimension, equal to zero at LPA.  

Writing the regulator as $R_k(\q)=k^2 \tilde R((\q/k)^2)$, the flows become 
\begin{equation}
\begin{split}
k\partial_k \tilde U_{k}=&-d \tilde U_{k}+(d-2)\tilde\rho\tilde U_{k}'
%\\&
+ \frac{1}{2 (kL)^d}\sum_{\tilde\q}(2\tilde R-\tilde\q^2 \tilde R' )\Bigg(\frac1{\tilde\q^2+\tilde R+\tilde U'_{k}+2\tilde\rho \tilde U_k''}
%\\&
+\frac{N-1}{\tilde\q^2+\tilde R(\tilde\q)+\tilde U'_{k}}\Bigg),
\end{split}
\label{eq_LPA_dimless}
\end{equation}
and
\begin{equation}
\begin{split}
k\partial_k \tilde I_{k}=&-d \tilde I_{k}+(d-2)\tcrhos\tilde I_{k}'
%\\&
+ \frac{1}{2 (kL)^d}\sum_{\tilde\q\neq0}(2\tilde R-\tilde\q^2 \tilde R' )\Bigg(\frac1{\tilde\q^2+\tilde R+\tilde I'_{k}+2\tcrhos \tilde I_k''}
%\\&
+\frac{N-1}{\tilde\q^2+\tilde R(\tilde\q)+\tilde I'_{k}}\Bigg),
\end{split}
\label{eq_LPA_dimless2}
\end{equation}

In the thermodynamic limit, both flow equations become the standard dimensionless LPA equation
\begin{equation}
k\partial_k \tilde U_{k}=-d \tilde U_{k}+(d-2)\tilde \rho\tilde U_{k}'+ \frac{1}{2}\int_{\tilde \q}(2\tilde R-\tilde\q^2 \tilde R' )\Bigg(\frac1{\tilde\q^2+\tilde R+\tilde U'_{k}+2\tilde \rho\tilde U''_{k}}+\frac{N-1}{\tilde\q^2+\tilde R+\tilde U'_{k}}\Bigg),
\label{eq_LPA_thermo}
\end{equation} 
and
\begin{equation}
k\partial_k \tilde I_{k}=-d \tilde I_{k}+(d-2)\tilde \tcrhos I_{k}'+ \frac{1}{2}\int_{\tilde \q}(2\tilde R-\tilde\q^2 \tilde R' )\Bigg(\frac1{\tilde\q^2+\tilde R+\tilde I'_{k}+2\tcrhos\tilde I''_{k}}+\frac{N-1}{\tilde\q^2+\tilde R+\tilde I'_{k}}\Bigg),
\label{eq_LPA_thermo2}
\end{equation} 
and at criticality, the dimensionless effective potential reaches a fixed point as $k\to 0$, that is, $\lim_{k\to 0}\lim_{L\to \infty}\tilde U_{k}=\lim_{k\to 0}\lim_{L\to \infty}\tilde I_{k}=\tilde U^*$, with $\tilde U^*$ the solution of
\begin{equation}
0=-d \tilde U^*+(d-2)\tilde\rho\tilde U^{*'}+  \frac{1}{2}\int_{\tilde \q}(2\tilde R-\tilde\q^2 \tilde R' )\Bigg(\frac{1}{\tilde\q^2+\tilde R+\tilde U^{*'}+2\tilde \rho\tilde U^{*''}}+\frac{N-1}{\tilde\q^2+\tilde R+\tilde U^{*'}}\Bigg).
\label{eq-FP}
\end{equation} 
It is important for the following to note that $\tilde U^*$, considered as a function of the modulus of the field $\tilde{\phi}=\sqrt{2\tilde\rho}$, is a non-convex function at small field (for $d<4$) whereas at $k=0$, the dimensionful effective potential $U_{k=0}$ is convex for all values of the field, as it should for a function obtained from a (true) Legendre transform.

Beyond LPA, $\eta$ takes a finite value at the fixed point, which, in particular, has the effect of turning $(d-2)$ factors in the second terms of all equations between Eqs.~\eqref{eq_LPA_dimless} and \eqref{eq-FP} into $(d-2+\eta)$. %At large field, the term involving an integral in Eq.~\eqref{eq-FP} can be neglected because either $\tilde U^{*'}$ diverges (in $2<d<4$) or vanishes (in $d\ge4$). 
For $2<d<4$, one expects the potential to increase at large fields. Thus, the integrand in Eq.~\eqref{eq-FP} becomes negligible in this regime.
This leads to a power law behavior of $\tilde U^*$
\begin{equation}
\tilde U^*\propto \tilde\rho^{\frac{d}{d-2+\eta}}.
\end{equation} 
The same power law behavior in the universal large field regime will be inherited by the rate function, giving 
\begin{equation}
\label{ratef_tail}
\tilde I\propto \tcrhos^{\frac{d}{d-2+\eta}},
\end{equation} 
 with $\eta=0$ at LPA. 

At finite size $L$, the flow equations \eqref{eq_LPA_dimless} and \eqref{eq_LPA_dimless2} are indistinguishable from Eqs.~\eqref{eq_LPA_thermo} and \eqref{eq_LPA_thermo2} as long as $k L\gg 1$. Therefore, at criticality ($T=T_c$), and for $a^{-1}\gg k\gg L^{-1}$,  both $\tilde U_k$ and $\tilde I_{k}$ flow towards the same fixed point solution $\tilde U^*$, i.e. $U_k(\rho)\simeq k^{d}\tilde U^*(k^{-(d-2)}\rho)$  and $I_k(\crho)\simeq U_k(\rho=\crho)\simeq k^{d}\tilde U^*(k^{-(d-2)}\crho)$, which corresponds to the correct scaling with exponent $d-2+\eta$ (with $\eta=0$ at LPA).

However, for $kL\lesssim 1$ the flows of $U_{k}$ and $I_{k}$ differ significantly. Indeed, the flow of the effective potential can be rewritten as
\begin{equation}
\partial_k U_{k}\simeq \frac{1}{2L^d}\partial_k R_{k}(0)\Bigg(\frac{1}{R_k(0)+U'_{k}+2\rho U''_{k}}+\frac{N-1}{R_k(0)+U'_{k}}\Bigg),
\end{equation} 
since the contribution of the finite momenta is negligible thanks to $\partial_k R_{k}(\q)$ in this regime.
This corresponds to the flow of a 0-dimensional field theory. Because $R_k(0)\to0$ as $k\to 0$, both $U'_{k}(\rho)$ and $U'_k(\rho)+2\rho U''_{k}(\rho)$ must become strictly positive as $k\to0$: This induces a return to convexity, with the minimum of the effective potential at $\rho=0$, which is expected for a (true) Legendre transform.
On the other hand, the flow of the rate function $I_k$ stops very quickly for $k\lesssim 1/L$ since the zero-momentum mode is absent from the sum and the other modes do not contribute.
Finally, at large field, the flows of $U_{k}$ and $I_{k}$ is barely modified by the finite size effects,~\footnote{See however \cite{Balog2024} for a discussion of the log-correction that appears in the rate function at large field.}  and we recover the powerlaw behavior of the effective potential and rate function, $U_{k=0}, I_{k=0}\propto \rho^{\frac{d}{d-2+\eta}}$ (with $\eta=0$ at LPA).

\subsection{Numerical integration of the LPA flow equation}
\label{sec_comparisons_numeric_int}

We are interested in computing the rate function $I_\zeta$ in the critical regime, which takes a universal form. 
It is therefore not necessary to take into account the early stage of the flow (say for a given lattice geometry at scale $k=\Lambda$ with a specific initial condition) since all these microscopic details will be washed out when flowing towards the fixed point. When the flow is initialized on the critical surface, the fixed point is almost reached at an intermediary (non-universal) scale $k_*$ while it is only when $k$ is of the order of $L^{-1}$ that finite size effects enter the flow.
For this reason, to get rid of the early stage RG flow, we first integrate the flow from an initial condition on the critical surface, down to a sufficiently small scale $k_*$ such that the running dimensionless potential $\tilde U_{k_*}$ and rate function $\tilde I_{k_*}$ are (almost) identical to the fixed point potential $\tilde U^*$. In a second step, we take as new initial condition for the RG flow what was obtained at $k_*$, that is $\tilde U^*$. In practice, we numerically solve the fixed point equation Eq.~\eqref{eq-FP}, and use this solution as the starting point of the flow. All quantities are then measured in units of this new ``microscopic" length scale $k_*$, which is equivalent to setting $k_*=1$. This has the advantage that all irrelevant perturbations to the fixed point have been set to zero, ensuring there will not be corrections to scaling.

Thus, we initialize the flows of the effective potential and rate function by $U_{k_*}(\rho)=k_*^d \tilde U^*(k_*^{-d+2}\rho)$ and $I_{k_*}(\crho)=k_*^d \tilde U^*(k_*^{-d+2}\crho)$. Such an initial condition corresponds to $\zeta=0$ (since the system is critical for $L\to\infty$, as it never leaves the fixed point). To generate a finite $\zeta$, we slightly perturb the fixed point, e.g. by adding a small quadratic term to the fixed point potential $\tilde U^*$. As this perturbation is relevant, this induces a flow into the ordered or disordered phase (in the thermodynamic limit) and allows us to vary $\zeta$, see below.

In practice, the field dependence of functions $U$ and $I$ is discretized on a grid such that the position of the minimum of $\tilde U^*$ is between $1/15$ and  $1/5$ of the grid. We use between $200$ and $400$ points on the grid and discretize derivatives using central derivatives with seven points. The fixed point solution, used as an initial condition is obtained by solving Eq.~\eqref{eq-FP} using a variant of the Newton-Raphson method.
For the numerical integration of the flow equations, Eqs.~\eqref{flow_u_lpa} and \eqref{flow_i_lpa}, we use the Euler method with RG time steps of $10^{-5}-10^{-4}$ (integration with Runge-Kutta of order 4 gives almost identical results). 
We run the flow in terms of dimensionless quantities, i.e. integrate Eq.~\eqref{eq_LPA_dimless} and Eq.~\eqref{eq_LPA_dimless2}, until $kL\simeq 2-10$. Then we switch to dimensionful quantities, i.e. integrate Eq.~\eqref{flow_i_lpa}, and run the flow until termination. The sum over momenta is performed by summing over all momenta $\q$ such that $|\q|<6k$, beyond which the summands are negligible as they are cut by $\partial_k R_k$, excluding (including) $\q=0$ for the flow of $I_k$ ($U_k$).
For $kL\gtrsim 40$, we replace the sums over momenta by integrals (i.e. we use Eqs.~\eqref{eq_LPA_thermo} and \eqref{eq_LPA_thermo2}) and we have checked that the transition from integrals to sums is smooth when $kL\simeq 40$. By varying all parameters of our numerical integration of the flows, we have checked that our results are numerically converged up to high precision.

In the following, we have used the following families of regulator functions
\begin{equation}
\begin{split}
    R_{k,\exp}(\q) &= \alpha k^2 e^{-\q^2/k^2} \\
R_{k,\theta}(\q)&=\alpha (k^2-\q^2) \theta(k^2-\q^2) \\
R_{k,W}(\q)&=\alpha \frac{\q^2}{(\exp(\q^2/k^2)-1)},
\end{split}
\label{eq_Rk}
\end{equation}
that we refer to as ``exponential", ``$\theta$" and ``Wetterich" regulators respectively. In principle, physical quantities, and in particular universal scaling functions, should not depend on the regulator chosen. However, approximations such as the LPA induce a spurious dependence on the regulator.
The derivative expansion, the LPA being the lowest order, has been argued to be a convergent series when the regulator is appropriately chosen, as observed for the critical exponents \cite{Balog2019}. Note that the $\theta$ regulator does not properly regularize the derivative expansion flow equations beyond order two, but works at LPA.
For a given regulator family, this convergence occurs when the optimal prefactor $\alpha$ is found at each order of the expansion so that the critical exponents reach an extremum. Here we chose to optimize the critical exponent $\nu$, see also \cite{DePolsi2022} for a discussion of the convergence of the optimized $\alpha$ with the order of the derivative expansion.

Studying the variations of various quantities (e.g. critical exponents) as a function of the regulator shape at a given approximation level allows for evaluating the stability of the method at that order. However, as detailed in \cite{depolsi2020}, the error bar estimated from these variations is typically too small and not a good estimate of the true error bar. To estimate the global error accurately, one would need to compute results at least two consecutive orders of the derivative expansion, which we are currently undertaking \cite{Rose2031}.

To compute the rate function for different $\zeta$, we first need to determine a mapping between the initial condition of the flow (here a quadratic perturbation of the fixed point)  and $\zeta$. In the disordered phase, the correlation length in the thermodynamic limit is given at LPA by \footnote{This definition corresponds to the so-called second-moment correlation length, the most easily estimated in Monte-Carlo simulations, which can be computed as $\xi^2= \partial_{\mathbf{p}^2}\big|_{\mathbf{p}=0} \Gamma^{(2)}(\mathbf{p};\rho=0)/\Gamma^{(2)}(\mathbf{p}=0;\rho=0)$, with $\partial_{\mathbf{p}^2}\big|_{\mathbf{p}=0} \Gamma^{(2)}(\mathbf{p};\rho=0)=1$ at LPA.}
\begin{equation}
    \xi_{\infty,{\rm LPA}}^2=\lim_{L\to\infty}\frac{1}{U'_{k=0}(\rho=0)}. 
    \label{eq_xiLPA}
\end{equation}
In practice, we integrate the flow of the effective potential in the thermodynamic limit, Eq.~\eqref{eq_LPA_thermo} with $M=0$, starting from the fixed point potential $\tilde U^*$ perturbed by a quadratic term $\delta r\tilde{\rho}$. To a very high level of precision, the system is critical for $\delta r=0$, while it flows to the disordered (ordered) phase for $\delta r>0$ ($\delta r<0$). For small enough $\delta r>0$, the correlation length behaves as
\begin{equation}
\label{eq_xi_frg}
    \xi_{\infty,{\rm LPA}}(\delta r)= \xi_{+,{\rm LPA}} \delta r^{-\nu_{\rm LPA}}.
\end{equation}
Here $\nu_{\rm LPA}$ is the critical exponent $\nu$ determined at LPA and $\xi_{+,{\rm LPA}}$ is a non-universal amplitude that we fit. We have computed them for the regulators \eqref{eq_Rk} at optimal $\alpha$ (optimized over $\nu_{\rm LPA}$ for each regulator), as summarized in Table \ref{tab:LPA_xi}. For comparison, the best available results for the critical exponents $\nu$ of the $O(N)$ models are found from FRG at the fourth order of derivative expansion \cite{depolsi2020} and they read $0.6716(6)$ for $N=2$ and $0.7114(9)$ for $N=3$. Then, for a given value of $L$ and $\zeta$, we choose $\delta r$ using Eq.~\eqref{eq_xi_frg} and the definition of $\zeta$,
\begin{equation}
\label{eq_zeta_FRG}
    \zeta={\rm sign}(\delta r) \frac{L}{\xi_{\infty,{\rm LPA}}(|\delta r|)},
\end{equation}
that is,
\begin{equation}
\label{eq_zeta_FRG}
    \delta r={\rm sign}(\zeta) \left(\frac{|\zeta| \xi_{+,{\rm LPA}}}L\right)^{1/\nu_{\rm LPA}}.
\end{equation}

\begin{table}
    \centering
%    \begin{tabular}{c|c|c|c}
%       reg.  & $\alpha_{opt}$ & $\xi_{+,{\rm LPA}}$ & $\nu_{\rm LPA}$ \\
%    \hline
%    \hline
%         \multicolumn{4}{c}{$N=2$}\\
%    \hline
%       $\theta$  & $1.0$ & $1.0444$ & $0.7082$ \\
%    \hline
%       exp.  & $4.65$ & $1.3029$ & $0.7106$ \\
%    \hline
%       Wett.  & $6.05$ & $1.4148$ & $0.7098$ \\
%    \hline
%         \multicolumn{4}{c}{$N=3$}\\
%    \hline
%       $\theta$  & $1.0$ & $1.0458$ & $0.7611$ \\
%    \hline
%       exp.  & $4.65$ & $1.3794$ & $0.7639$ \\
%    \hline
%        Wett. & $6.05$ & $1.5297$ & $0.7631$ \\
%    \hline
%    \end{tabular}
   \begin{tabular}{c|c|c|c|c} $N$ & reg. & $\alpha_{opt}$ & $\xi_{+,{\rm LPA}}$ & $\nu_{\rm LPA}$ \\ \hline \hline   & $\theta$ & $1.0$ & $1.0444$ & $0.7082$ \\ $2$ & exp. & $4.65$ & $1.3029$ & $0.7106$ \\   & Wett. & $6.05$ & $1.4148$ & $0.7098$ \\ \hline  & $\theta$ & $1.0$ & $1.0458$ & $0.7611$ \\ $3$ & exp. & $4.65$ & $1.3794$ & $0.7639$ \\  & Wett. & $6.05$ & $1.5297$ & $0.7631$ \\ \hline \end{tabular}
    \caption{Numerical parameters from the FRG for the Eq. \eqref{eq_xi_frg} at LPA.}
    \label{tab:LPA_xi}
\end{table}

%Recently a method was proposed \cite{ihssen2023,batini23} for discretizing the FRG flow equations which greatly stabilizes the flow. In present work we have applied such discretization procedure as well as a more standard one involving central derivatives through a number of points (typically we use 7 points). We find that in the parameter ranges that we study here, there is no significant difference in the results for the rate function obtained by one method or another. What we have noticed however is that the standard central derivatives method tends to break while the method from Wink etal. is much more stable in 2 cases. First when trying to capture the tail behavior (by manipulating the cutoff $k$ at which the switch to dimensionful quantities is done), the standard method works only up to some value of dimensionless argument. Secondly the standard method tends to break down if the grid mesh $\Delta\tilde{\rho}$ is decreased beyond some small value while keeping the box range fixed. Note however that the results by the standard method stabilize completely before such a small $\Delta\tilde{\rho}$ is reached when studying the critical behavior. 

\subsection{Rate function at large $N$ from FRG}

\label{sec_comparisons_benchm_Ninf}

Let us consider the large $N$ limit, where the LPA flow of the potential becomes exact, as a benchmark for the numerical resolution of the flow equation. In this limit, starting from Eq.~\eqref{eq_LPA_thermo2}, we can derive the gap equation \eqref{eq_gap_rate_proto} straightforwardly. Following the reasoning from \cite{Tetradis1996}, we first take the limit $N\to\infty$ in Eq.~\eqref{flow_i_lpa} to obtain the corresponding large $N$ flow equation.
\begin{equation}
\label{eq_flow_I_FRG_Ninf}
    \partial_k I_k(\crho)=\frac{N}{2L^d}\sum_{\q\neq 0}\frac{\partial_k R_{k}(\q)}{\q^2+R_k(\q)+I'_k(\crho)}.
\end{equation}

Now, we define $\check{f}_{k}(\cW)=\crho$ as the inverse function of $\cW_k(\crho)\equiv I_k'(\crho)$. Note that since $\check \Gamma_{k=\Lambda}[\bphi]=\mathcal H[\bphi]$, we have that $I_{k=\Lambda}(\crho)= V(\crho)$ and thus $\check f_{k=\Lambda}(\cW)=f_{\Lambda}(\cW)$, with $f_{\Lambda}$ defined in Sec.~\ref{sec_eff_pot_Ninf}.

One readily finds the flow equation of $\check{f}_{k}$,
\begin{equation}
    \partial_k \check{f}_{k}(\cW)=\frac{N}{2L^d}\sum_{\q\neq 0}\frac{\partial_k R_{k}(\q)}{(\q^2+R_k(\q)+\cW)^2},
\end{equation}
 the right-hand side of which is a total derivative with respect to $k$.
Integrating this equation in $k$ from $0$ to $\Lambda$ and using that $\lim_{k\to 0}R_k=0$ and $\lim_{k\to \Lambda}R_k=\infty$, we recover the gap equation \eqref{eq_gap_rate_proto} for the rate function,
\begin{equation}
    \crho=f_{\Lambda}(\cW)-\frac{N}{2L^d}\sum_{\q\neq 0}\frac{1}{\q^2+\cW},
\end{equation}
from which the derivation detailed in Sec. \ref{sec_rate_Ninf} follows. Clearly, any regulator dependence disappears from the end result.

The same is not true however for the fixed point potential. One can show that the fixed point solution depends on the regulator at $N\to\infty$ \cite{knorr2021}, where the LPA is exact. The corresponding fixed-point potential equation reads
\begin{equation}
- 2\tilde U_*' +(d-2)\trho\tilde U_*''-\tilde U_*''N L_1(\tilde U_*')=0,
\label{eq_FP}
\end{equation}
where the threshold function $L_1(w)$ depends on the regulator $R_k(\q)=k^2 \tilde R(\q^2/k^2)$,
\begin{equation}
L_1(w)=\frac12 \int_{\tilde \q}\frac{2\tilde R(\tilde\q^2)-\tilde\q^2 \tilde R'(\tilde\q^2)  }{\left(\tilde \q^2+\tilde R(\tilde\q^2)+w\right)^2}.
\label{eq_threshold}
\end{equation}
Introducing $\tilde W(\trho)=\tilde U_*' (\trho)$, it has been shown in \cite{knorr2021} that the solution to Eq.~\eqref{eq_FP} reads for an arbitrary regulator
\begin{equation}
\trho =\tilde F_d\left(\tilde W(\trho);\tilde R\right)\equiv \frac{N}{2(d-2)}\int_{\tilde \q}\frac{2\tilde R(\tilde\q^2)-\tilde\q^2 \tilde R'(\tilde\q^2) }{\left(\tilde \q^2+\tilde R(\tilde\q^2)\right)^2} {}_2F_1\left(2,1-\frac d2;2 -\frac d2\bigg|-\frac{\tilde W(\trho)}{\tilde \q^2+\tilde R(\tilde\q^2)}\right),
\label{eq_FPsol}
\end{equation}
where ${}_2F_1(a,b,c|-z)$ is a hypergeometric function.

Note that $\tilde F_d(w;\tilde R)$ has a shape qualitatively similar to that of $\cF_d(w)$, see Eq.~\eqref{eq_def_cF}.
Indeed, the singularity of the hypergeometric function at $z=-1$, ${}_2F_1\left(2,1-\frac d2;2 -\frac d2\bigg|-z\right)\simeq \frac{2-d}{2(z+1)}$ corresponds to the singularity at $w=w_{-\infty}\equiv-\min_{\tilde \q}( \tilde \q^2+\tilde R(\tilde\q^2))$ in Eq.~\eqref{eq_threshold}, similar to the pole at $z=-\pi$ of $\cF_d(z)$, reached formally for $\trho\to-\infty$.
On the other hand, in the limit $\trho\to\infty$, i.e. $W(\trho)\to\infty$, Eq.~\eqref{eq_FP} gives $W(\trho)\sim \trho^{\frac2{d-2}}$, which is recovered from Eq.~\eqref{eq_FPsol} using that ${}_2F_1\left(2,1-\frac d2;2 -\frac d2\bigg|-z\right)\sim z^{\frac{d-2}2}$ for $z\to\infty$, as does $\cF_d(z)$. Finally, $\tilde F_d(w;\tilde R)=0$ for some $w_0<0$. This explains why the fixed point potential $\tilde U^*$ and the rate function have similar shapes, although the former is regulator dependent while the latter is universal (but depends on the boundary conditions). 

While the shape of $\tilde U^*$ depends on the regulator, the critical exponents, relevant or irrelevant, are independent of it \cite{knorr2021}. By analyzing the stability matrix of the LPA flow equation, we find numerically in $d=3$ that $\nu=1/(d-2)$ up to numerical uncertainty at the level of $5$-th digit, independently of the regulator shape or prefactor $\alpha$.

\begin{figure}[!t]
    \centering
    \includegraphics[width=8cm]{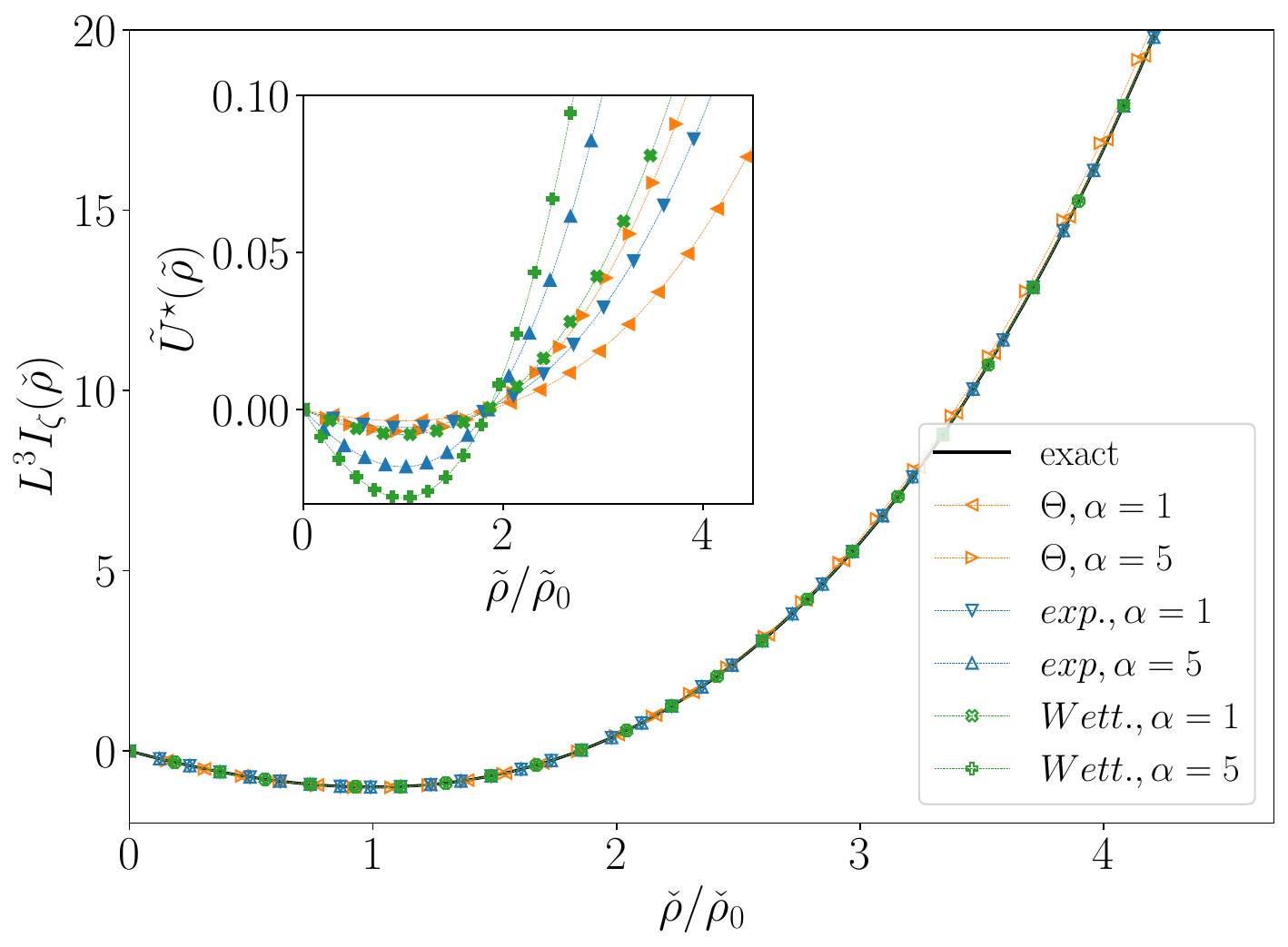}
    \caption{
    Rate functions at $\zeta=0$ in the large $N$ limit in $d=3$ obtained from FRG with different regulators, compared with the exact result. The inset shows the initial conditions of the flows (corresponding to the regulator-dependent fixed points). }
    \label{fig_Rdependence_Ninf}
\end{figure}

Although we can compute the rate function exactly in the present case, we nevertheless examine now the FRG results obtained by numerical integration of the LPA flow equation (as explained in Sec.~\ref{sec_comparisons_numeric_int}) to prepare for the next section. Figure \ref{fig_Rdependence_Ninf} shows rate functions for different regulators compared with the exact result. Despite widely varying initial conditions, the rate functions converge to the exact result with high numerical precision by the end of the flow.

An interesting quantity is the ratio between the probability of the average magnetization (corresponding to $\crho=0$) and the probability of the most probable total spin (corresponding to a $\crho_0$). If the rate function has a non-trivial minimum (corresponding to $\crho_0>0$), this ratio is related to the difference between $I_{min,N}\equiv I(\crho_0)$ and $I_{0,N}\equiv I(0)$. Since the rate function scales as $L^{-d}$ in the scaling regime and is proportional to $N$ at large $N$, it is convenient to introduce the universal amplitude  $\Delta I_N=L^d(I_{min,N}-I_{0,N})/N$, which depends on $\zeta$.~\footnote{\label{foot:PRL} This corrects a statement of Ref.~\cite{Balog22}, where we wrote that there was a non-universal amplitude associated with the scale of the rate function. }
 
 The exact value of $\Delta I_N$ at $\zeta=0$ in large $N$ is obtained from Eq.~\eqref{eq_gap_rate},
 \begin{equation}
 \begin{split}
     \Delta I_\infty&=\frac{L^d}{N}\int_0^{\crho_0} \cW(\crho)d\crho,\\
                    &=-4\pi\int^{0}_{\check z_0} \cF_d(z) dz,
 \end{split}
\end{equation}
where we used $\crho = NL^{2-d} \cF_d(z)$, $z=L^2 \cW/4\pi$ and we recall that $\check z_0$ is such that $\cF(\check z_0)=0$. In three dimensions, it reads $\Delta I_\infty\sim -1.00028...$. Table \ref{tab:imin} summarizes the comparison of $\Delta I_\infty$ between this exact result and that obtained by integrating numerically in the same dimension the LPA equation.

\begin{table}
    \centering
    \begin{tabular}{c |c}
         & $\Delta I_\infty$ \\
    \hline
    \hline
     exact  & $-1.00028211..$ \\
    \hline
      $\theta$, $\alpha=1$   & $-1.00319$ \\
      $\theta$, $\alpha=5$  & $-1.01548$ \\
      exp., $\alpha=1$  & $-1.00013$ \\
      exp., $\alpha=5$  & $-1.00004$\\
      Wett. $\alpha=1$   & $-0.99983$\\
      Wett. $\alpha=5$   & $-1.00005$ \\
    \end{tabular}
    \caption{Universal amplitude $\Delta I_\infty$ at $\zeta=0$, see text, in the large $N$ limit obtained from different regulators and compared to the exact result. %The specific numerical parameters used in those calculations are $\delta\rho=0.001$, $n=400$ points and $L=10^4$. We have used the Euler integration with $dl=-1\cdot 10^{-4}$ but checked that the Runge Kutta method of order 4 gives the same result up to roughly four-five digits. 
    }
    \label{tab:imin}
\end{table}

\begin{figure}
    \centering
    \includegraphics[width=8cm]{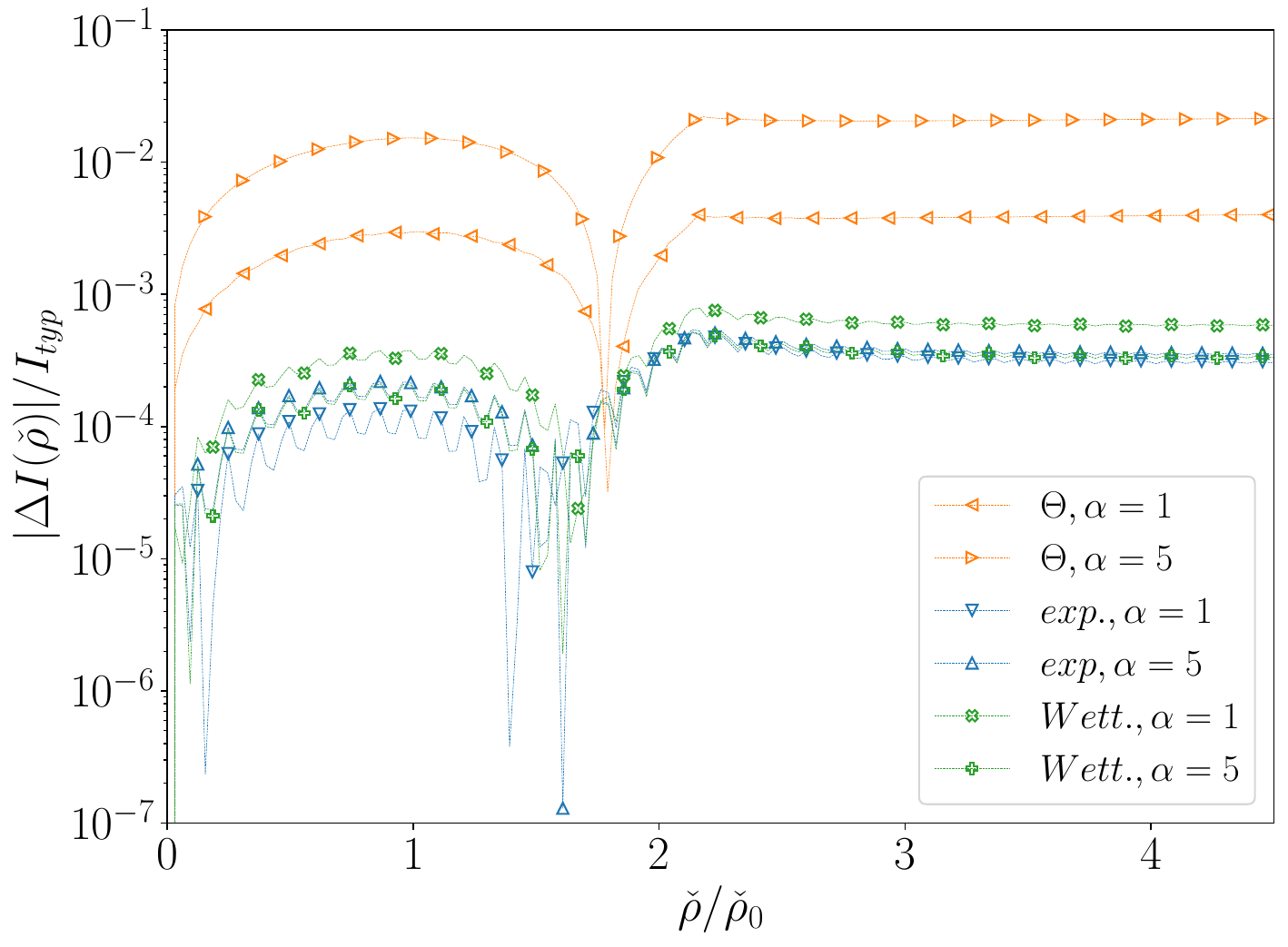}
    \caption{
    Relative error of the rate function obtained from numerically solving LPA flow equation in the large $N$ limit with respect to the exact result, for various regulators. Here $\Delta I=I_{num}-I_{exact}$ and we compare it to $I_{typ}(\crho)={\rm max}(I_{exact}(\crho),1)$, since the rate function is of order one and negative and changes sign for $\crho\simeq 2\crho_0$.}
    \label{fig:rate_error_Ninfty}
\end{figure}

In Fig. \ref{fig:rate_error_Ninfty}, we show the relative error between the exact and FRG results for different regulators. For analytic regulators, the error is about $10^{-4}$. In the case of the $\theta$ regulator, the increased error is likely due to its nonanalytic properties, which affect the accuracy of the flow. Indeed the flow equation changes abruptly each time $k$ changes by $2\pi/L$ since $\partial_k R_k(\q)\propto \theta(k^2-\q^2)$ in this case, see also \cite{Fister2015}.
These results demonstrate the high numerical accuracy of our numerical integration of the LPA flow equation, giving confidence in our numerical integration of the flows at finite $N$.

%To summarize, our workflow to compute the rate function for various $\zeta_{FRG}$ for a given $L$ (to compare to MC simulations we use $L=10000$) is: i) Solve the flow of $U_k$ in the thermodynamic limit, tuning $\delta$ such that $\xi_\infty=L/\zeta_{FRG}$; ii) Solve the flow of the rate function (at finite $L$) with the same $\delta$ if $\zeta_{FRG}>0$ or with $-\delta$ if $\zeta_{FRG}<0$.

\section{Rate functions at finite $N$ in dimension three}
\label{sec_comparisons}

For finite $N$, an exact solution for the rate function is not available, and the LPA provides only an approximation to the exact theory. In addition to the expected regulator dependence of the fixed point functions, some spurious regulator dependence remains in the rate functions after the flow is integrated, as explained in Sec.~\ref{sec_comparisons_numeric_int}. Since the variance in FRG results caused by changing the regulator prefactor is not informative, we will only use the optimal prefactors given in Table~\ref{tab:LPA_xi} for $N=2$ and $N=3$. We compare the FRG results at LPA with Monte Carlo (MC) results, which when $L$ is large enough to ensure convergence, serve as the closest available proxy to exact results. From now on, we only consider three-dimensional systems.

\begin{figure}
    \centering
    \includegraphics[width=8cm]{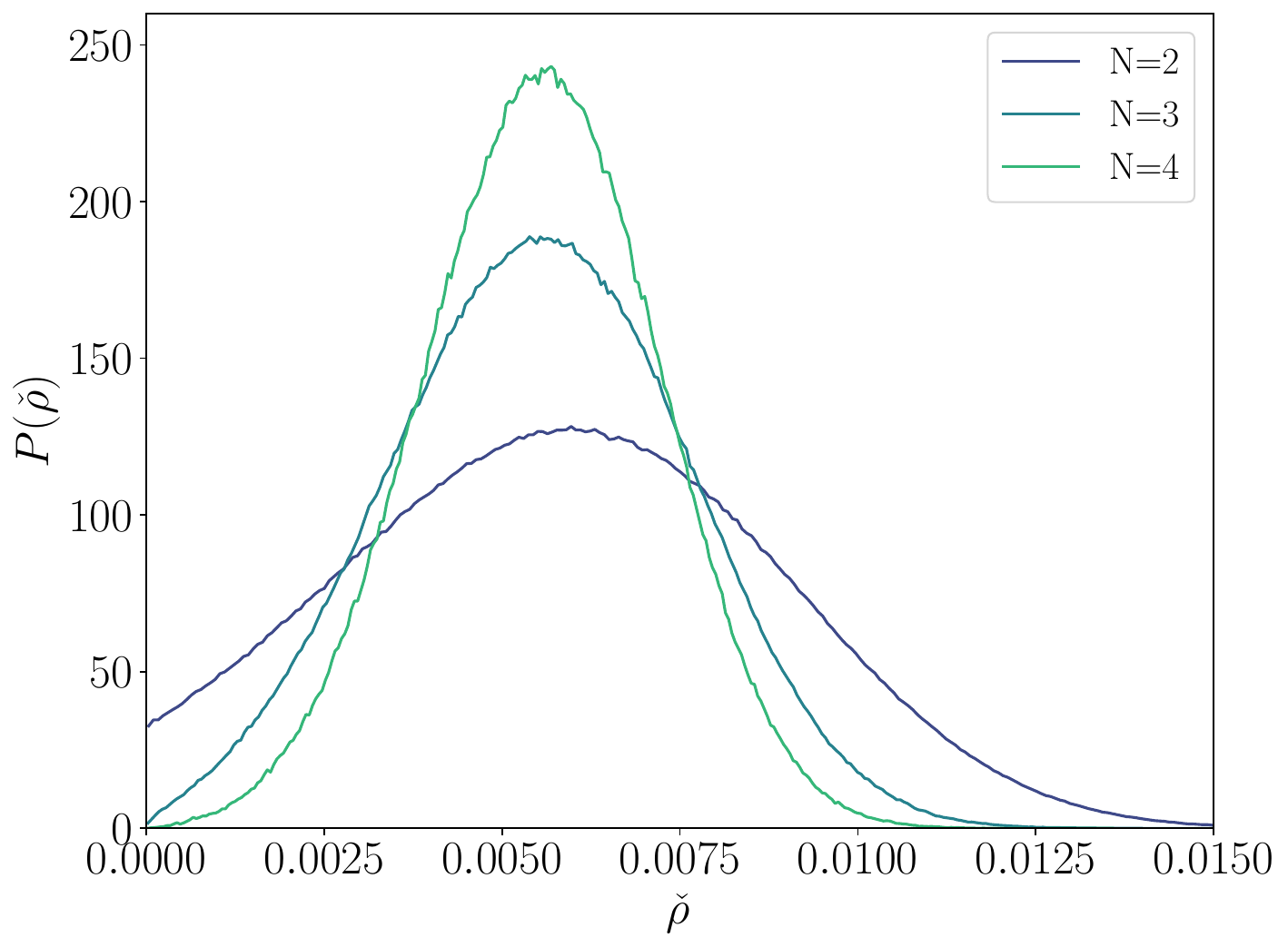}
    \caption{
      Probability distributions $\langle \delta(\hat \s^2/2-\crho)\rangle$ as a function of $\crho$ for different $N$ at $\zeta=0$ computed from MC at $L=96$. The Jacobian term $\propto \crho^{(N-2)/2}$ has not been extracted from it yet to obtain $P(\s)$}
    \label{fig_PDF_diffN_zet0}
\end{figure}

Our MC simulations of the 3D $O(2)$ (XY),  $O(3)$ (Heisenberg) and $O(4)$ models were performed using Wolff cluster algorithm \cite{Wolff1989} with typically $10^7$ independent MC configurations. We do not directly evaluate histograms at given $T$ but record for each MC configuration the squared total spin density $\hat \s^2/2$ and the energy density  $L^{-3}\sum_{\langle i,j\rangle}\hat \bs_i.\hat \bs_j$, where $\hat \bs_i$ are unit-length $O(N)$ spins and the sum is over nearest-neighbors. This allows us to use histogram reweighting techniques \cite{newman1999monte} to extend the calculation of the rate function to a range of $\zeta$.
Our results show that the difference between the scaled rate function at $L=96$ and $L=128$ is negligible for both $N=2$ and $N=3$. The PDF so-obtained at $\zeta=0$ are shown in Fig.~\ref{fig_PDF_diffN_zet0}. Note that by recording the squared magnetization we are in effect computing $\langle \delta(\hat \s^2/2-\crho)\rangle$, which differs from $P(\s)$ by a Jacobian term $\propto \crho^{(N-2)/2}$. We therefore divide the former by the Jacobian to reconstruct the rate functions.

To vary $\zeta$, we need the critical temperatures and correlation length dependence on it for our specific models, as both are non-universal. Fortunately, these are available in the literature, and we used 
\begin{equation}
    \xi_{\infty,{\rm MC}} = \xi_{+,{\rm MC}}\left(1-\frac{T_c}{T}\right)^{-\nu},
\end{equation}
with $J/T_c\simeq 0.45416$  and $\xi_{+,{\rm MC}}\simeq 0.498$ for $N=2$  ($J/T_c\simeq 0.6930$  and $\xi_{+}\simeq 0.484$ for $N=3$)  \cite{butera1998,hasenbusch2008,Deng2005}.
To compute the PDF for various $L$ and fixed $\zeta$,  the temperature is therefore chosen such that $\xi_{\infty,{\rm MC}}=L/\zeta$. 
%\ar{I think we have complicated the analysis for nothing in the  PRL, the correlation length we get from FRG is in fact the second-moment correlation length (and not the one of the exponential decay), since $\xi_{2nd}^2\equiv \int_\x \x^2 G(\x)/2d \int_\x G(\x)=-\partial_{\q^2}G(\q)|_{\q=0}/G(\q=0)$ and this is exactly $\partial_{\q^2}\Gamma^{(2)}(\q=0)/U''=\xi_{FRG}^2$. So no need to translate $\xi_{2nd}$ of MC to $\xi_{exp}$. Even though we ``corrected'' for it in the PRL, it won't change anything (since the ratio is so close to 1). Should we also correct that in the future erratum? }

Finally, let us note that the typical scale of the field is non-universal, meaning that there is a non-universal amplitude associated with $\crho$ to be fixed. Since in all cases considered here, the rate function has a non-trivial minimum at $\zeta=0$, we use the position of this minimum at $\zeta=0$, called $\crho_0$ to normalize the $\crho$. This  allows for a direct comparison between MC simulations and FRG calculation. 

% It was demonstrated by calculations that the ratio between $\frac{\xi_{exp.}}{\xi_{2nd}}\approx 1.0002$ for the N=2 case when $T>T_C$ \cite{campostrini2001}, however a general argument exists that the ratio is always very close to $1$ when $T>T_C$ \cite{caselle2017}. So when considering a system with $T>T_C$, even in the critical region we shall always assume the $\xi_{2nd}\equiv \xi_{exp.}$.

\subsection{Comparison between FRG and MC}

\label{sec_comparisons_N_MC}

\begin{figure}
    \centering
    \includegraphics[width=8cm]{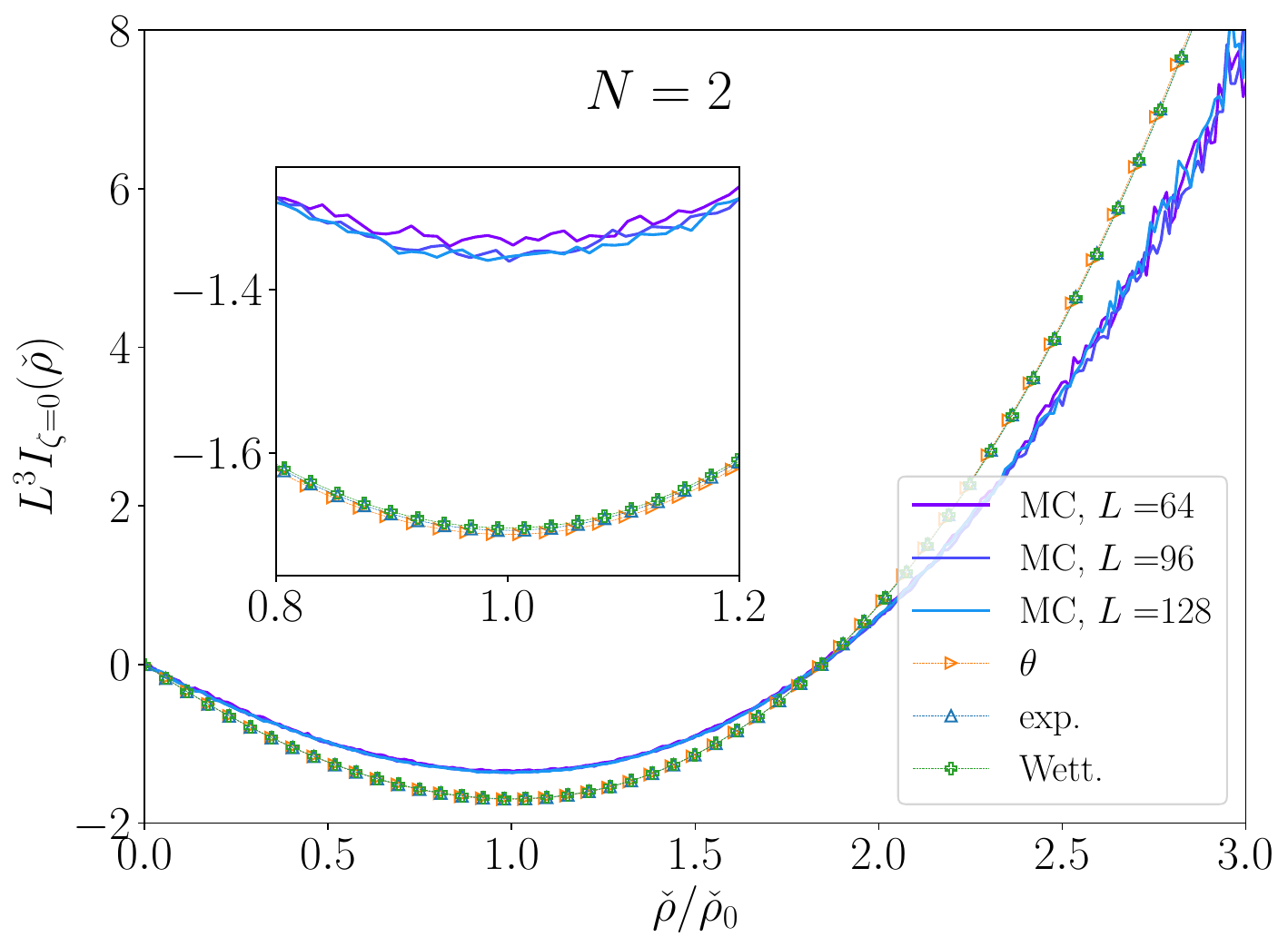}
        \caption{Rate functions of the $O(2)$ model at $\zeta=0$ obtained from MC for various system sizes and  FRG at LPA for various regulators at optimal $\alpha$. The inset zooms near the minimum.}
    \label{fig:I_n2}
\end{figure}

\begin{figure}
    \centering
    \includegraphics[width=8cm]{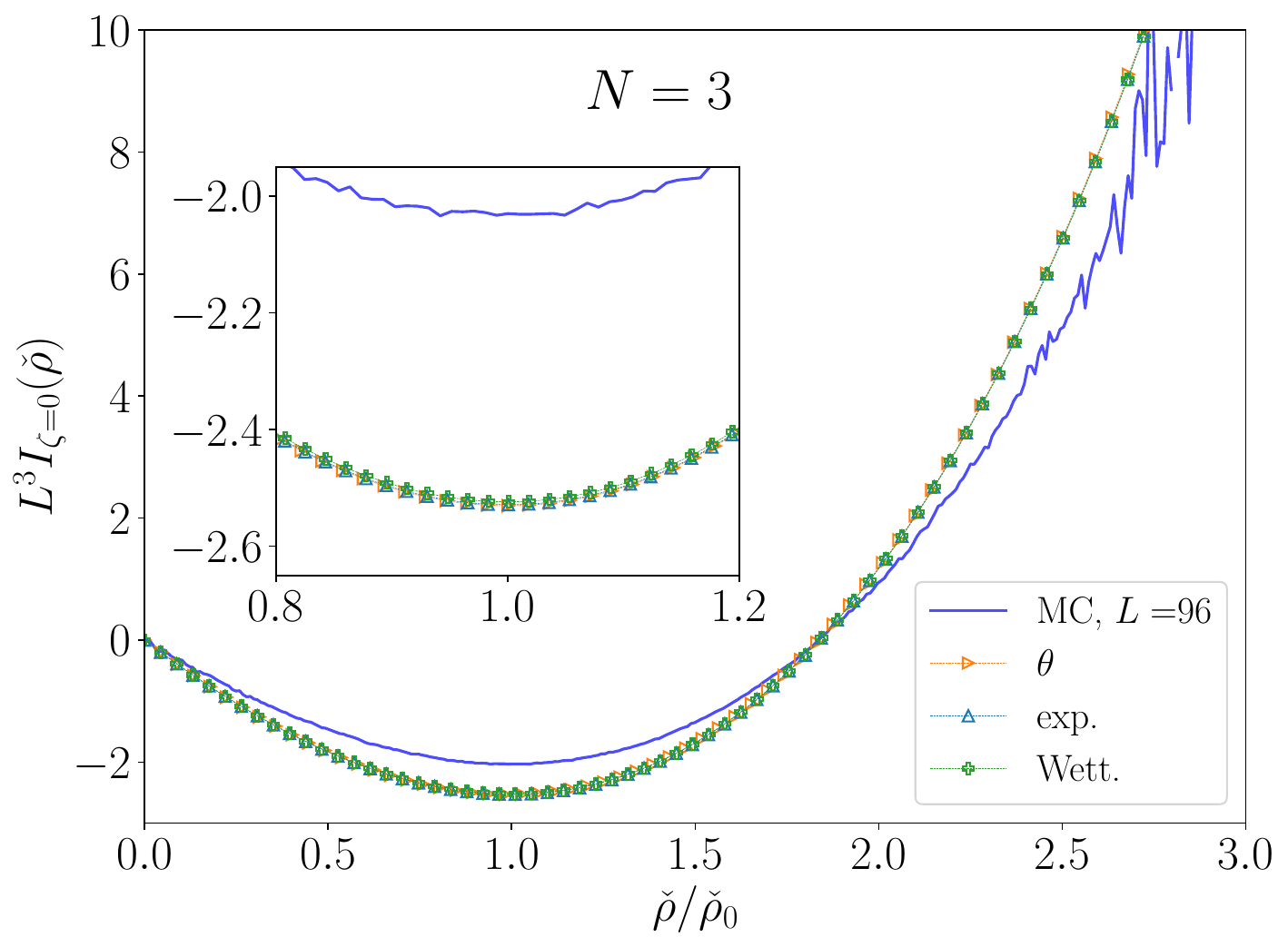}
    \caption{
    Rate functions of the $O(3)$ model at $\zeta=0$ obtained from and  FRG at LPA for various regulators at optimal $\alpha$. The inset zooms near the minimum.}
    \label{fig:I_n3}
\end{figure}

\begin{figure}
    \centering
    \includegraphics[width=16cm]{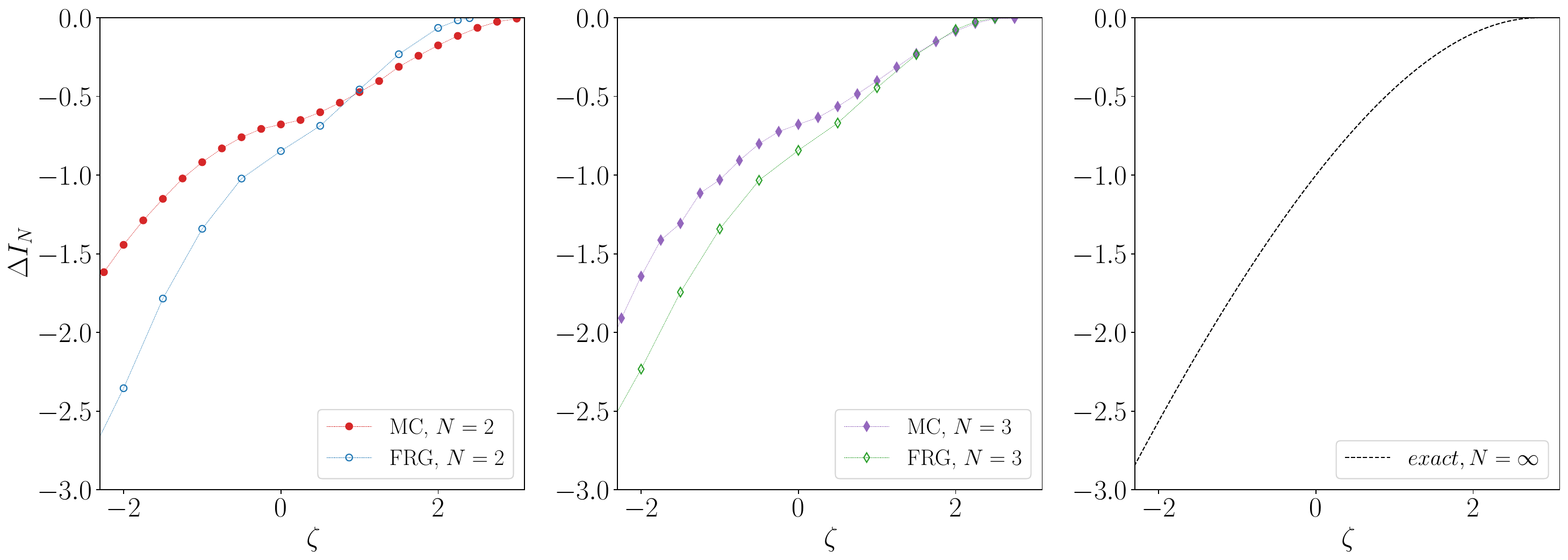}
    \caption{
    Universal amplitude  $\Delta I_N=\frac{L^d(I_{min,N}-I_{0,N})}{N}$ as a function of $\zeta$ for different $N$, obtained from FRG and MC and from the exact solution in large $N$.}
    \label{fig_imin_zeta_all}
\end{figure}

\begin{table}[]
    \centering
\begin{tabular}{c | c | c | c | c} & \multicolumn{2}{c|}{$N=2$} & \multicolumn{2}{c}{$N=3$} \\ & $\zeta_c$ & $N\Delta I_N$ & $\zeta_c$ & $N\Delta I_N$ \\ \hline \hline MC & $3.2\pm 0.2$ & $-1.36\pm 0.01$ & $2.7\pm 0.2$ & $-2.03\pm 0.01$ \\ \hline $\theta$ & $2.4931$ & $-1.6998$ & $2.5839$ & $-2.5306$ \\ exp. & $2.4904$ & $-1.6948$ & $2.5810$ & $-2.5289$ \\ Wett. & $2.4887$ & $-1.6919$ & $2.5795$ & $-2.5238$ \\ \end{tabular}
    \caption{Universal amplitude $\Delta I_N=\frac{L^d(I_{min,N}-I_{0,N})}{N}$ at $\zeta=0$  and critical ratio  $\zeta_c$  obtained from FRG for different regulators at optimal $\alpha$ compared to MC. }
    \label{tab:Imin_zetac_Nfinite}
\end{table}

Figures \ref{fig:I_n2} and \ref{fig:I_n3} compare the rate functions at $\zeta=0$ for $N=2$ and $3$, respectively, as obtained from MC and FRG. We observe that while the shapes are similar, there is a discrepancy between the FRG result at LPA and the MC data. This cannot be due to finite-size corrections in the latter, as the variation from $L=64$ to $L=128$ is much smaller than the difference with FRG. Furthermore, we find a very small variation of the FRG results when changing regulators, as exemplified in Table \ref{tab:Imin_zetac_Nfinite}, which gives the universal amplitude $\Delta I_N=\frac{L^d(I_{min,N}-I_{0,N})}{N}$.
This is clearly an issue due to the LPA, which is not exact for $N=2$ and $N=3$.

So far, we have focused on the case $\zeta = 0$. To test our method's ability to capture universality, we examine the behavior of the rate functions for different values of $\zeta$, as these should all be universal. We observe that the dependence of the rate functions on $\zeta$ for all $N$ is qualitatively similar to the $N\to \infty$ case shown in Fig.~\ref{fig:I_crit_Ninf}. A concise way to characterize this universality is by examining the dependence of $\Delta I_N$ on $\zeta$ for different $N$. The combined results are shown in Fig.~\ref{fig_imin_zeta_all}. The FRG results for $N\to \infty$ are excluded, as they match the exact results to high precision, as discussed in Sec.~\ref{sec_comparisons_benchm_Ninf}.

As in the large $N$ limit, we find that for finite $N$, the rate functions change convexity at a critical value $\zeta_c>0$ and consequently $\Delta I_N\to 0$, as shown in Fig.~\ref{fig_imin_zeta_all}. In FRG calculations, $I_\zeta'(\crho)$ is computed directly during the flow, and $\zeta_c$ is most easily identified as the point where $I_\zeta'(\crho = 0)$ crosses zero. For MC, although $\Delta I_N$ can be determined with high precision when it is not much smaller than 1, we find it more accurate to estimate $\zeta_c$ by calculating $I_\zeta'(\crho = 0)$ using histogram reweighting for different values of $\zeta$, as this quantity depends linearly on $\zeta$ near $\zeta_c$. This critical value  $\zeta_c$ is given in Table \ref{tab:Imin_zetac_Nfinite} for $N = 2$ and $N = 3$, and we find that the discrepancies between MC and FRG at LPA are approximately $25\%$, similar to the results for $\Delta I_N$.

%When calculating the rate functions by the FRG method at LPA for the $O(N)$ model with $N=2$ and $3$, we find qualitatively correct behavior of the rate function with the confidence of the method, devised from variance induced by the regulator shape changes, being quite high at $< 0.5\%$. However the qualitative error of assessment of the universal numbers characterizing rate functions can be as high as $\approx 25\%$. This is quite surprising since in both these cases the anomalous dimension of the field $\eta$ is quite small, implying $LPA$ should be a good approximation and furthermore the error in the exponents $\nu$ at LPA is between $5-8\%$ from the exact values. This leaves much to be desired in determination of the rate functions and work is underway to improve the approximation level and see how it affects the accuracy. 

\subsection{Shapes of the rate functions}
\label{sec_comparisons_shapes}

\begin{figure}
    \centering
    \includegraphics[width=8cm]{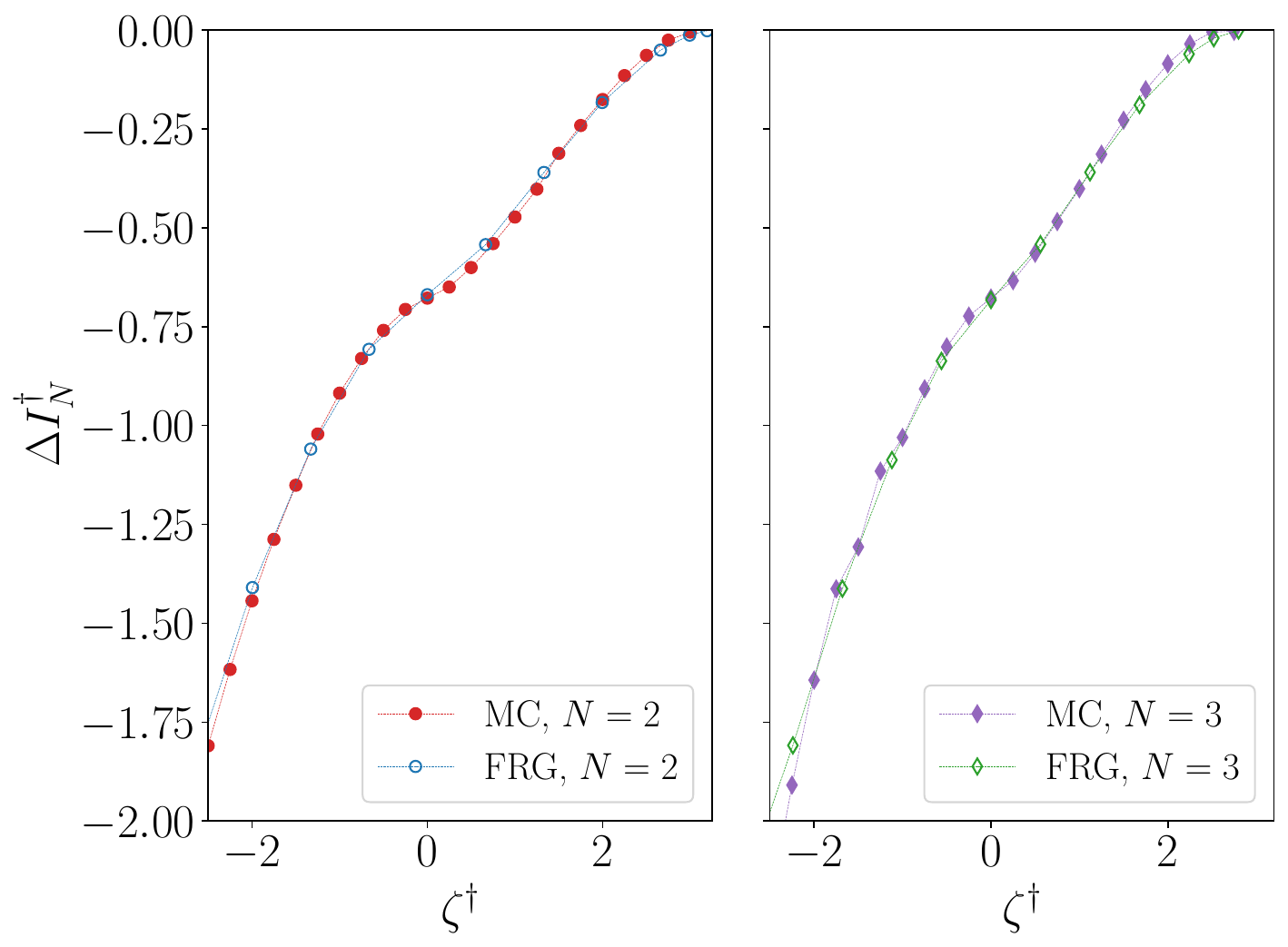}
    \caption{
    Universal amplitude $\Delta I_N^\dag=\frac{L^d(I^{\dagger}_{min,N}-I^{\dagger}_{0,N})}{N}$ as a function of  $\zeta^{\dagger}$ for $N=2$ and $3$. For MC data, $\zeta^{\dagger}=\zeta$ and $I^{\dagger}(\crho)=I(\crho)$ while for FRG $\zeta^{\dagger}=r_{\zeta}(N)\zeta $ and $I^{\dagger}(\crho)= r_{I}(N)I(\crho)$, see text.}
    \label{fig_imin_zeta_coll}
\end{figure}

While we have seen that there is quite a discrepancy between FRG at LPA and MC simulations, we now show that the shape of the rate function is very well captured at LPA, if we correct for the amplitude of the rate function (i.e. allow ourselves to multiply it by a factor $r_I(N)$) and of $\zeta$ (i.e. allow ourselves to multiply it by a factor $r_\zeta(N)$). Those factors are found by finding the best collapse between FRG and MC data of $\Delta I_N$ as a function of $\zeta$, as shown in Fig.~\ref{fig_imin_zeta_coll}, giving $r_{I}(2)\approx 0.79$ and $r_{\zeta}(2)\approx 1.33$ for $N=2$ and $r_{I}(3)\approx 0.81$ and $r_{\zeta}(3)\approx 1.12$ for $N=3$.~\footnote{In comparing FRG at LPA and MC for $N=1$ in \cite{Balog22}, we had allowed ourselves to rescale the amplitude of the rate function, although for an incorrect reason, see footnote \ref{foot:PRL}. This amounted to use $r_I(1)\approx 0.80$ and $r_\zeta(1)\simeq 1.11$. }

\begin{figure}
    \centering
    \includegraphics[width=8cm]{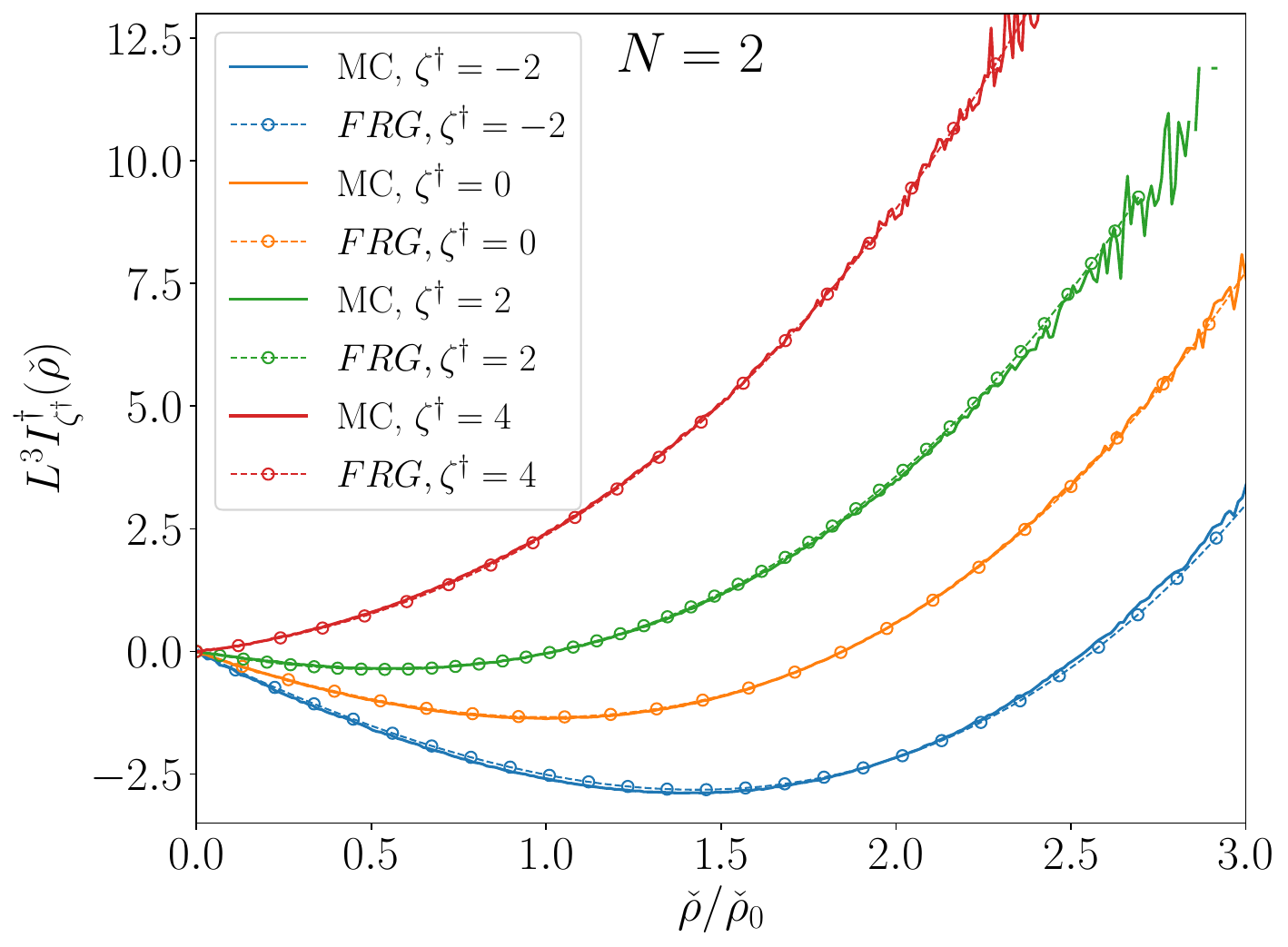}
    \caption{
    Rate functions as functions of $\crho$ calculated by MC and FRG (with exponential regulator) for various $\zeta^{\dagger}$ for the $O(2)$ model. For MC data, $\zeta^{\dagger}=\zeta$ and $I^{\dagger}(\crho)=I(\crho)$ while for FRG $\zeta^{\dagger}=r_{\zeta}(N)\zeta $ and $I^{\dagger}(\crho)= r_{I}(N)I(\crho)$, see text. }
    \label{fig_idagger_coll}
\end{figure}

Focusing on the case $N=2$, for which there is the most discrepancy, Fig.~\ref{fig_idagger_coll} shows the comparison of rate functions for various $\zeta$ between MC and FRG, once we have accounted for a rescaling of $\zeta$ and amplitude of the rate function. We then find an excellent functional collapse both in field and in $\zeta$. The relative error between the FRG and MC rate functions, as a function of $\crho$, is less than $1\%$. This suggests that the error induced by the LPA is primarily concentrated in the calculation of the two universal constants $\Delta I_N$ and $\zeta_c$, while the finer details of the rate function's dependence on $\crho$ and $\zeta$ are reproduced with high accuracy. This has also been observed in the calculation of the rate function of the 3D Ising model using FRG \cite{Balog22} and perturbative RG \cite{Sahu2024}.

\begin{figure}
    \centering
    \includegraphics[width=8cm]{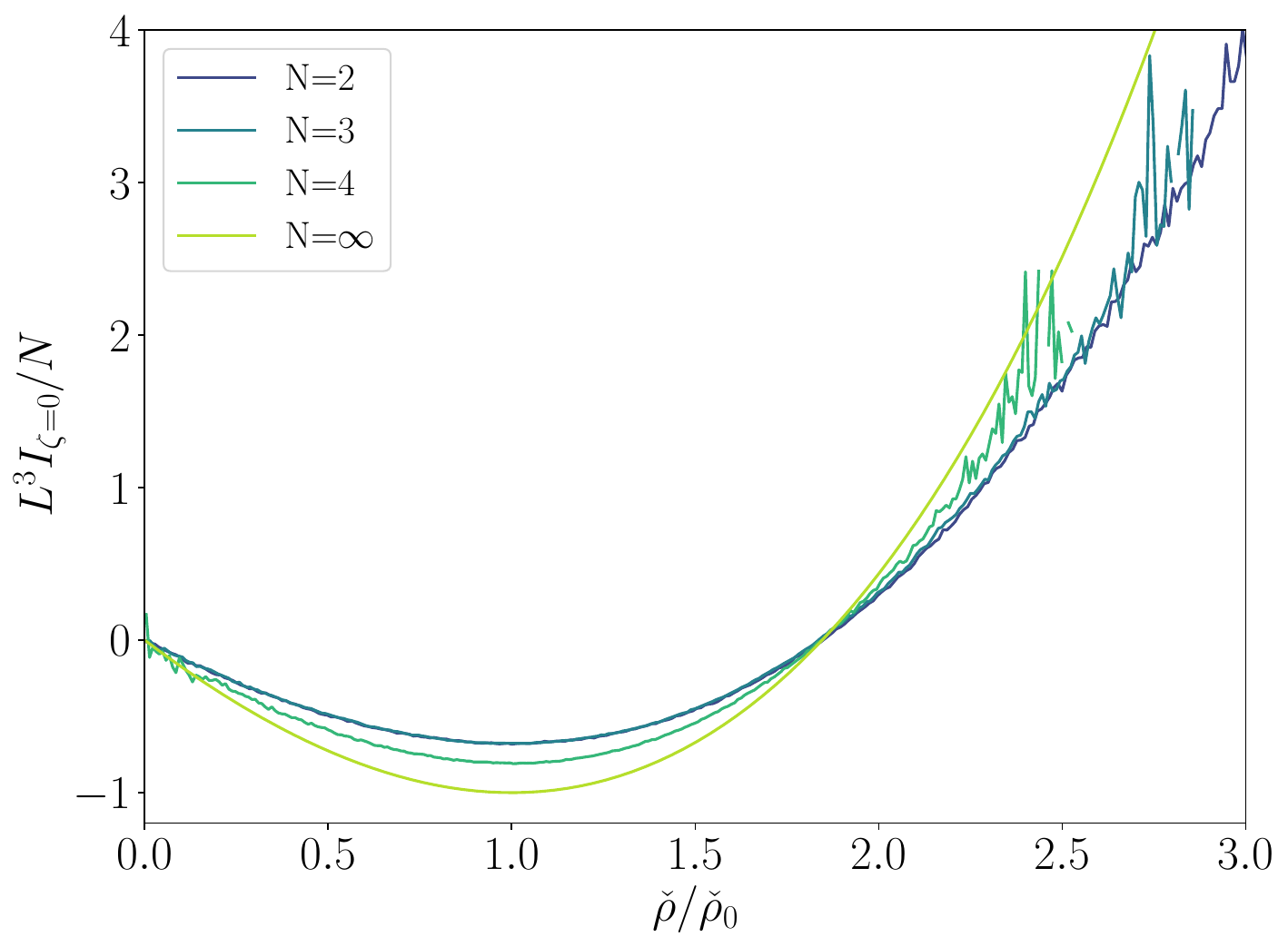}
    \caption{
            Rate functions for different $N$ at $\zeta=0$, obtained from MC at $L=96$ for finite $N$, and from the exact result for large $N$. Surprisingly, the $N=2$ and $N=3$ rate functions are almost on top of each other when rescaled by $N$.  }
    \label{fig_diffN_zet0}
\end{figure}

Finally, we compare in Fig.~\ref{fig_diffN_zet0} the shape of the rate functions at $\zeta=0$ for various $N$ as obtained from MC and in the large $N$ limit. Since the rate function is of order $N$, we compare the rate functions normalized by $N$. Very surprisingly, we observe that the normalized rate functions at $N=2$ and $N=3$ are almost identical, while the rate function for $N=4$ is closer to that of large $N$.~\footnote{We used $T_c/J=1.0684(1)$  for $N=4$ \cite{Kanaya1995}.} We leave a more detailed understanding of this for future work.

\section{Conclusion}
\label{sec_discussion}
We have conducted a detailed study of the rate function in the $O(N)$ models. In the  $N \to \infty$ limit, we derived an exact solution, similar to the well-known derivation of the effective potential in this limit \cite{Moshe2003}. This result provides valuable insights into the general $O(N)$ case, as it qualitatively captures the key behaviors across the full range of $N$.

We then applied the nonperturbative approach to calculating the rate function, first introduced in \cite{Balog22}, to the $O(N)$ models. This approach leverages the full power of the FRG, allowing us to handle cases where exact solutions are not available. We focused on testing the reliability of results obtained using the Local Potential Approximation (LPA), the lowest order of the derivative expansion, which is the most widely used approximation in FRG studies.

By benchmarking against the $N \to \infty$ case, where the LPA becomes exact, we confirmed the robustness of our numerical implementation of the method. For finite $N$, the dependence on the infrared regulator introduces no more than a $0.5\%$ variation. However, estimating the universal scales characterizing the rate functions proved more challenging, with discrepancies of up to $25\%$. Despite this, the nontrivial universal features of the rate functions are qualitatively well captured, even at the LPA level. 

Looking forward, future work will focus on improving the precision of universal scale estimates by extending the approximation level beyond LPA. Additionally, we aim to generalize the method to other models, in particular disordered systems and out-of-equilibrium statistical physics.

\acknowledgements

IB wishes to thank Martin Hasenbusch for his valuable correspondence. AR and BD thank the Institute of Physics in Zagreb for its hospitality, and BD and IB for that of the Universit\'e de Lille, where parts of this work were done.
AR and IB are supported by the Croatian Science fund project HRZZ-IP-10-2022-9423, an IEA CNRS project, and by the “PHC COGITO” program (project number: 49149VE) funded by the French Ministry for Europe and Foreign Affairs, the French Ministry for Higher Education and Research, and The Croatian Ministry of Science and Education. IB wishes to acknowledge the support of the INFaR and FrustKor projects financed by the EU through the National Recovery and Resilience Plan (NRRP) 2021-2026.

% We thank  O. B\'enichou, N. Dupuis, G. Tarjus and  N. Wschebor for discussions and feedbacks. BD thanks F. Benitez, M. Tissier and Z. R\'acz for many  discussions in an early stage of this work. AR also thanks G. Verley for discussions on this and related subjects. IB acknowledges the support of the the QuantiXLie Centre of Excellence, a project cofinanced by the Croatian Government and European Union through the European Regional Development Fund - the Competitiveness and Cohesion Operational Programme (Grant KK.01.1.1.01.0004). BD acknowledges the support from the French ANR through the project NeqFluids (grant ANR-18-CE92-0019).
%AR is supported by the Research Grants QRITiC I-SITE ULNE/ ANR-16-IDEX-0004 ULNE.

\bibliography{CLT_paper}

%merlin.mbs apsrev4-1.bst 2010-07-25 4.21a (PWD, AO, DPC) hacked
%Control: key (0)
%Control: author (72) initials jnrlst
%Control: editor formatted (1) identically to author
%Control: production of article title (-1) disabled
%Control: page (0) single
%Control: year (1) truncated
%Control: production of eprint (0) enabled
\begin{thebibliography}{67}%
\makeatletter
\providecommand \@ifxundefined [1]{%
 \@ifx{#1\undefined}
}%
\providecommand \@ifnum [1]{%
 \ifnum #1\expandafter \@firstoftwo
 \else \expandafter \@secondoftwo
 \fi
}%
\providecommand \@ifx [1]{%
 \ifx #1\expandafter \@firstoftwo
 \else \expandafter \@secondoftwo
 \fi
}%
\providecommand \natexlab [1]{#1}%
\providecommand \enquote  [1]{``#1''}%
\providecommand \bibnamefont  [1]{#1}%
\providecommand \bibfnamefont [1]{#1}%
\providecommand \citenamefont [1]{#1}%
\providecommand \href@noop [0]{\@secondoftwo}%
\providecommand \href [0]{\begingroup \@sanitize@url \@href}%
\providecommand \@href[1]{\@@startlink{#1}\@@href}%
\providecommand \@@href[1]{\endgroup#1\@@endlink}%
\providecommand \@sanitize@url [0]{\catcode `\\12\catcode `\$12\catcode
  `\&12\catcode `\#12\catcode `\^12\catcode `\_12\catcode `\%12\relax}%
\providecommand \@@startlink[1]{}%
\providecommand \@@endlink[0]{}%
\providecommand \url  [0]{\begingroup\@sanitize@url \@url }%
\providecommand \@url [1]{\endgroup\@href {#1}{\urlprefix }}%
\providecommand \urlprefix  [0]{URL }%
\providecommand \Eprint [0]{\href }%
\providecommand \doibase [0]{http://dx.doi.org/}%
\providecommand \selectlanguage [0]{\@gobble}%
\providecommand \bibinfo  [0]{\@secondoftwo}%
\providecommand \bibfield  [0]{\@secondoftwo}%
\providecommand \translation [1]{[#1]}%
\providecommand \BibitemOpen [0]{}%
\providecommand \bibitemStop [0]{}%
\providecommand \bibitemNoStop [0]{.\EOS\space}%
\providecommand \EOS [0]{\spacefactor3000\relax}%
\providecommand \BibitemShut  [1]{\csname bibitem#1\endcsname}%
\let\auto@bib@innerbib\@empty
%</preamble>
\bibitem [{\citenamefont {Fukuda}\ and\ \citenamefont
  {Kyriakopoulos}(1975)}]{Fukuda1975}%
  \BibitemOpen
  \bibfield  {author} {\bibinfo {author} {\bibfnamefont {R.}~\bibnamefont
  {Fukuda}}\ and\ \bibinfo {author} {\bibfnamefont {E.}~\bibnamefont
  {Kyriakopoulos}},\ }\href {\doibase
  https://doi.org/10.1016/0550-3213(75)90014-0} {\bibfield  {journal} {\bibinfo
   {journal} {Nuclear Physics B}\ }\textbf {\bibinfo {volume} {85}},\ \bibinfo
  {pages} {354} (\bibinfo {year} {1975})}\BibitemShut {NoStop}%
\bibitem [{\citenamefont {O'Raifeartaigh}\ \emph {et~al.}(1986)\citenamefont
  {O'Raifeartaigh}, \citenamefont {Wipf},\ and\ \citenamefont
  {Yoneyama}}]{ORaifeartaigh1986}%
  \BibitemOpen
  \bibfield  {author} {\bibinfo {author} {\bibfnamefont {L.}~\bibnamefont
  {O'Raifeartaigh}}, \bibinfo {author} {\bibfnamefont {A.}~\bibnamefont
  {Wipf}}, \ and\ \bibinfo {author} {\bibfnamefont {H.}~\bibnamefont
  {Yoneyama}},\ }\href {\doibase https://doi.org/10.1016/S0550-3213(86)80031-1}
  {\bibfield  {journal} {\bibinfo  {journal} {Nuclear Physics B}\ }\textbf
  {\bibinfo {volume} {271}},\ \bibinfo {pages} {653} (\bibinfo {year}
  {1986})}\BibitemShut {NoStop}%
\bibitem [{\citenamefont {Mukaida}\ and\ \citenamefont
  {Shimada}(1996)}]{Mukaida1996}%
  \BibitemOpen
  \bibfield  {author} {\bibinfo {author} {\bibfnamefont {H.}~\bibnamefont
  {Mukaida}}\ and\ \bibinfo {author} {\bibfnamefont {Y.}~\bibnamefont
  {Shimada}},\ }\href {\doibase https://doi.org/10.1016/0550-3213(96)00403-8}
  {\bibfield  {journal} {\bibinfo  {journal} {Nuclear Physics B}\ }\textbf
  {\bibinfo {volume} {479}},\ \bibinfo {pages} {663} (\bibinfo {year}
  {1996})}\BibitemShut {NoStop}%
\bibitem [{\citenamefont {Bouchaud}\ and\ \citenamefont
  {Georges}(1990)}]{Bouchaud1990}%
  \BibitemOpen
  \bibfield  {author} {\bibinfo {author} {\bibfnamefont {J.-P.}\ \bibnamefont
  {Bouchaud}}\ and\ \bibinfo {author} {\bibfnamefont {A.}~\bibnamefont
  {Georges}},\ }\href {\doibase https://doi.org/10.1016/0370-1573(90)90099-N}
  {\bibfield  {journal} {\bibinfo  {journal} {Physics Reports}\ }\textbf
  {\bibinfo {volume} {195}},\ \bibinfo {pages} {127} (\bibinfo {year}
  {1990})}\BibitemShut {NoStop}%
\bibitem [{\citenamefont {Privman}\ and\ \citenamefont
  {Fisher}(1984)}]{privman1984}%
  \BibitemOpen
  \bibfield  {author} {\bibinfo {author} {\bibfnamefont {V.}~\bibnamefont
  {Privman}}\ and\ \bibinfo {author} {\bibfnamefont {M.~E.}\ \bibnamefont
  {Fisher}},\ }\href {\doibase 10.1103/PhysRevB.30.322} {\bibfield  {journal}
  {\bibinfo  {journal} {Phys. Rev. B}\ }\textbf {\bibinfo {volume} {30}},\
  \bibinfo {pages} {322} (\bibinfo {year} {1984})}\BibitemShut {NoStop}%
\bibitem [{\citenamefont {Binder}(1981)}]{Binder1981}%
  \BibitemOpen
  \bibfield  {author} {\bibinfo {author} {\bibfnamefont {K.}~\bibnamefont
  {Binder}},\ }\href {\doibase 10.1103/PhysRevLett.47.693} {\bibfield
  {journal} {\bibinfo  {journal} {Phys. Rev. Lett.}\ }\textbf {\bibinfo
  {volume} {47}},\ \bibinfo {pages} {693} (\bibinfo {year} {1981})}\BibitemShut
  {NoStop}%
\bibitem [{\citenamefont {Bruce}\ and\ \citenamefont
  {Wilding}(1992)}]{Bruce1992}%
  \BibitemOpen
  \bibfield  {author} {\bibinfo {author} {\bibfnamefont {A.~D.}\ \bibnamefont
  {Bruce}}\ and\ \bibinfo {author} {\bibfnamefont {N.~B.}\ \bibnamefont
  {Wilding}},\ }\href {\doibase 10.1103/PhysRevLett.68.193} {\bibfield
  {journal} {\bibinfo  {journal} {Phys. Rev. Lett.}\ }\textbf {\bibinfo
  {volume} {68}},\ \bibinfo {pages} {193} (\bibinfo {year} {1992})}\BibitemShut
  {NoStop}%
\bibitem [{\citenamefont {Tsypin}(1994)}]{Tsypin1994}%
  \BibitemOpen
  \bibfield  {author} {\bibinfo {author} {\bibfnamefont {M.~M.}\ \bibnamefont
  {Tsypin}},\ }\href {\doibase 10.1103/PhysRevLett.73.2015} {\bibfield
  {journal} {\bibinfo  {journal} {Phys. Rev. Lett.}\ }\textbf {\bibinfo
  {volume} {73}},\ \bibinfo {pages} {2015} (\bibinfo {year}
  {1994})}\BibitemShut {NoStop}%
\bibitem [{\citenamefont {Tsypin}\ and\ \citenamefont
  {Bl\"ote}(2000)}]{Tsypin2000}%
  \BibitemOpen
  \bibfield  {author} {\bibinfo {author} {\bibfnamefont {M.~M.}\ \bibnamefont
  {Tsypin}}\ and\ \bibinfo {author} {\bibfnamefont {H.~W.~J.}\ \bibnamefont
  {Bl\"ote}},\ }\href {\doibase 10.1103/PhysRevE.62.73} {\bibfield  {journal}
  {\bibinfo  {journal} {Phys. Rev. E}\ }\textbf {\bibinfo {volume} {62}},\
  \bibinfo {pages} {73} (\bibinfo {year} {2000})}\BibitemShut {NoStop}%
\bibitem [{\citenamefont {Hilfer}\ \emph {et~al.}(2003)\citenamefont {Hilfer},
  \citenamefont {Biswal}, \citenamefont {Mattutis},\ and\ \citenamefont
  {Janke}}]{Hilfer2003}%
  \BibitemOpen
  \bibfield  {author} {\bibinfo {author} {\bibfnamefont {R.}~\bibnamefont
  {Hilfer}}, \bibinfo {author} {\bibfnamefont {B.}~\bibnamefont {Biswal}},
  \bibinfo {author} {\bibfnamefont {H.~G.}\ \bibnamefont {Mattutis}}, \ and\
  \bibinfo {author} {\bibfnamefont {W.}~\bibnamefont {Janke}},\ }\href
  {\doibase 10.1103/PhysRevE.68.046123} {\bibfield  {journal} {\bibinfo
  {journal} {Phys. Rev. E}\ }\textbf {\bibinfo {volume} {68}},\ \bibinfo
  {pages} {046123} (\bibinfo {year} {2003})}\BibitemShut {NoStop}%
\bibitem [{\citenamefont {Hilfer}\ \emph {et~al.}(2005)\citenamefont {Hilfer},
  \citenamefont {Biswal}, \citenamefont {Mattutis},\ and\ \citenamefont
  {Janke}}]{Hilfer2005}%
  \BibitemOpen
  \bibfield  {author} {\bibinfo {author} {\bibfnamefont {R.}~\bibnamefont
  {Hilfer}}, \bibinfo {author} {\bibfnamefont {B.}~\bibnamefont {Biswal}},
  \bibinfo {author} {\bibfnamefont {H.-G.}\ \bibnamefont {Mattutis}}, \ and\
  \bibinfo {author} {\bibfnamefont {W.}~\bibnamefont {Janke}},\ }\href
  {\doibase https://doi.org/10.1016/j.cpc.2005.03.053} {\bibfield  {journal}
  {\bibinfo  {journal} {Computer Physics Communications}\ }\textbf {\bibinfo
  {volume} {169}},\ \bibinfo {pages} {230} (\bibinfo {year}
  {2005})}\BibitemShut {NoStop}%
\bibitem [{\citenamefont {Malakis}\ and\ \citenamefont
  {Fytas}(2006)}]{Malakis2006}%
  \BibitemOpen
  \bibfield  {author} {\bibinfo {author} {\bibfnamefont {A.}~\bibnamefont
  {Malakis}}\ and\ \bibinfo {author} {\bibfnamefont {N.~G.}\ \bibnamefont
  {Fytas}},\ }\href {\doibase https://doi.org/10.1016/j.physa.2006.01.018}
  {\bibfield  {journal} {\bibinfo  {journal} {Physica A}\ }\textbf {\bibinfo
  {volume} {365}},\ \bibinfo {pages} {197} (\bibinfo {year}
  {2006})}\BibitemShut {NoStop}%
\bibitem [{\citenamefont {Xu}\ \emph {et~al.}(2020)\citenamefont {Xu},
  \citenamefont {Ferrenberg},\ and\ \citenamefont {Landau}}]{Xu2020}%
  \BibitemOpen
  \bibfield  {author} {\bibinfo {author} {\bibfnamefont {J.}~\bibnamefont
  {Xu}}, \bibinfo {author} {\bibfnamefont {A.~M.}\ \bibnamefont {Ferrenberg}},
  \ and\ \bibinfo {author} {\bibfnamefont {D.~P.}\ \bibnamefont {Landau}},\
  }\href {\doibase 10.1103/PhysRevE.101.023315} {\bibfield  {journal} {\bibinfo
   {journal} {Phys. Rev. E}\ }\textbf {\bibinfo {volume} {101}},\ \bibinfo
  {pages} {023315} (\bibinfo {year} {2020})}\BibitemShut {NoStop}%
\bibitem [{\citenamefont {Hasenfratz}\ \emph {et~al.}(1987)\citenamefont
  {Hasenfratz}, \citenamefont {Neuhaus}, \citenamefont {Jansen}, \citenamefont
  {Yoneyama},\ and\ \citenamefont {Lang}}]{Hasenfratz1987}%
  \BibitemOpen
  \bibfield  {author} {\bibinfo {author} {\bibfnamefont {A.}~\bibnamefont
  {Hasenfratz}}, \bibinfo {author} {\bibfnamefont {T.}~\bibnamefont {Neuhaus}},
  \bibinfo {author} {\bibfnamefont {K.}~\bibnamefont {Jansen}}, \bibinfo
  {author} {\bibfnamefont {H.}~\bibnamefont {Yoneyama}}, \ and\ \bibinfo
  {author} {\bibfnamefont {C.~B.}\ \bibnamefont {Lang}},\ }\href {\doibase
  https://doi.org/10.1016/0370-2693(87)91622-4} {\bibfield  {journal} {\bibinfo
   {journal} {Physics Letters B}\ }\textbf {\bibinfo {volume} {199}},\ \bibinfo
  {pages} {531} (\bibinfo {year} {1987})}\BibitemShut {NoStop}%
\bibitem [{\citenamefont {Kuti}\ and\ \citenamefont {Shen}(1988)}]{Kuti1988}%
  \BibitemOpen
  \bibfield  {author} {\bibinfo {author} {\bibfnamefont {J.}~\bibnamefont
  {Kuti}}\ and\ \bibinfo {author} {\bibfnamefont {Y.}~\bibnamefont {Shen}},\
  }\href@noop {} {\bibfield  {journal} {\bibinfo  {journal} {Physical Review
  Letters}\ }\textbf {\bibinfo {volume} {60}},\ \bibinfo {pages} {85} (\bibinfo
  {year} {1988})}\BibitemShut {NoStop}%
\bibitem [{\citenamefont {Hasenfratz}\ \emph {et~al.}(1989)\citenamefont
  {Hasenfratz}, \citenamefont {Jansen}, \citenamefont {Jersa´k}, \citenamefont
  {Lang}, \citenamefont {Neuhaus},\ and\ \citenamefont
  {Yoneyama}}]{Hasenfratz1989}%
  \BibitemOpen
  \bibfield  {author} {\bibinfo {author} {\bibfnamefont {A.}~\bibnamefont
  {Hasenfratz}}, \bibinfo {author} {\bibfnamefont {K.}~\bibnamefont {Jansen}},
  \bibinfo {author} {\bibfnamefont {J.}~\bibnamefont {Jersa´k}}, \bibinfo
  {author} {\bibfnamefont {C.}~\bibnamefont {Lang}}, \bibinfo {author}
  {\bibfnamefont {T.}~\bibnamefont {Neuhaus}}, \ and\ \bibinfo {author}
  {\bibfnamefont {H.}~\bibnamefont {Yoneyama}},\ }\href {\doibase
  https://doi.org/10.1016/0550-3213(89)90562-2} {\bibfield  {journal} {\bibinfo
   {journal} {Nuclear Physics B}\ }\textbf {\bibinfo {volume} {317}},\ \bibinfo
  {pages} {81} (\bibinfo {year} {1989})}\BibitemShut {NoStop}%
\bibitem [{\citenamefont {G\"ockeler}\ and\ \citenamefont
  {Leutwyler}(1991{\natexlab{a}})}]{Gockeler1991}%
  \BibitemOpen
  \bibfield  {author} {\bibinfo {author} {\bibfnamefont {M.}~\bibnamefont
  {G\"ockeler}}\ and\ \bibinfo {author} {\bibfnamefont {H.}~\bibnamefont
  {Leutwyler}},\ }\href {\doibase https://doi.org/10.1016/0550-3213(91)90246-T}
  {\bibfield  {journal} {\bibinfo  {journal} {Nuclear Physics B}\ }\textbf
  {\bibinfo {volume} {361}},\ \bibinfo {pages} {392} (\bibinfo {year}
  {1991}{\natexlab{a}})}\BibitemShut {NoStop}%
\bibitem [{\citenamefont {G\"ockeler}\ and\ \citenamefont
  {Leutwyler}(1991{\natexlab{b}})}]{Gockeler1991a}%
  \BibitemOpen
  \bibfield  {author} {\bibinfo {author} {\bibfnamefont {M.}~\bibnamefont
  {G\"ockeler}}\ and\ \bibinfo {author} {\bibfnamefont {H.}~\bibnamefont
  {Leutwyler}},\ }\href {\doibase https://doi.org/10.1016/0370-2693(91)91383-7}
  {\bibfield  {journal} {\bibinfo  {journal} {Physics Letters B}\ }\textbf
  {\bibinfo {volume} {253}},\ \bibinfo {pages} {193} (\bibinfo {year}
  {1991}{\natexlab{b}})}\BibitemShut {NoStop}%
\bibitem [{\citenamefont {G\"ockeler}\ and\ \citenamefont
  {Leutwyler}(1991{\natexlab{c}})}]{Gockeler1991b}%
  \BibitemOpen
  \bibfield  {author} {\bibinfo {author} {\bibfnamefont {M.}~\bibnamefont
  {G\"ockeler}}\ and\ \bibinfo {author} {\bibfnamefont {H.}~\bibnamefont
  {Leutwyler}},\ }\href {\doibase https://doi.org/10.1016/0550-3213(91)90260-5}
  {\bibfield  {journal} {\bibinfo  {journal} {Nuclear Physics B}\ }\textbf
  {\bibinfo {volume} {350}},\ \bibinfo {pages} {228} (\bibinfo {year}
  {1991}{\natexlab{c}})}\BibitemShut {NoStop}%
\bibitem [{\citenamefont {Dimitrovi\'c}\ \emph {et~al.}(1991)\citenamefont
  {Dimitrovi\'c}, \citenamefont {Nager}, \citenamefont {Jansen},\ and\
  \citenamefont {Neuhaus}}]{Dimitrovic1991}%
  \BibitemOpen
  \bibfield  {author} {\bibinfo {author} {\bibfnamefont {I.}~\bibnamefont
  {Dimitrovi\'c}}, \bibinfo {author} {\bibfnamefont {J.}~\bibnamefont {Nager}},
  \bibinfo {author} {\bibfnamefont {K.}~\bibnamefont {Jansen}}, \ and\ \bibinfo
  {author} {\bibfnamefont {T.}~\bibnamefont {Neuhaus}},\ }\href {\doibase
  https://doi.org/10.1016/0370-2693(91)91598-P} {\bibfield  {journal} {\bibinfo
   {journal} {Physics Letters B}\ }\textbf {\bibinfo {volume} {268}},\ \bibinfo
  {pages} {408} (\bibinfo {year} {1991})}\BibitemShut {NoStop}%
\bibitem [{\citenamefont {G\"ockeler}\ \emph {et~al.}(1993)\citenamefont
  {G\"ockeler}, \citenamefont {Kastrup}, \citenamefont {Neuhaus},\ and\
  \citenamefont {Zimmermann}}]{Gockeler1993}%
  \BibitemOpen
  \bibfield  {author} {\bibinfo {author} {\bibfnamefont {M.}~\bibnamefont
  {G\"ockeler}}, \bibinfo {author} {\bibfnamefont {H.~A.}\ \bibnamefont
  {Kastrup}}, \bibinfo {author} {\bibfnamefont {T.}~\bibnamefont {Neuhaus}}, \
  and\ \bibinfo {author} {\bibfnamefont {F.}~\bibnamefont {Zimmermann}},\
  }\href {\doibase https://doi.org/10.1016/0550-3213(93)90489-C} {\bibfield
  {journal} {\bibinfo  {journal} {Nuclear Physics B}\ }\textbf {\bibinfo
  {volume} {404}},\ \bibinfo {pages} {517} (\bibinfo {year}
  {1993})}\BibitemShut {NoStop}%
\bibitem [{\citenamefont {{Bramwell}}\ \emph {et~al.}(1998)\citenamefont
  {{Bramwell}}, \citenamefont {{Holdsworth}},\ and\ \citenamefont
  {{Pinton}}}]{BHP_1}%
  \BibitemOpen
  \bibfield  {author} {\bibinfo {author} {\bibfnamefont {S.~T.}\ \bibnamefont
  {{Bramwell}}}, \bibinfo {author} {\bibfnamefont {P.~C.~W.}\ \bibnamefont
  {{Holdsworth}}}, \ and\ \bibinfo {author} {\bibfnamefont {J.~F.}\
  \bibnamefont {{Pinton}}},\ }\href {\doibase 10.1038/25083} {\bibfield
  {journal} {\bibinfo  {journal} {\nat}\ }\textbf {\bibinfo {volume} {396}},\
  \bibinfo {pages} {552} (\bibinfo {year} {1998})}\BibitemShut {NoStop}%
\bibitem [{\citenamefont {{Bramwell}}\ \emph {et~al.}(2000)\citenamefont
  {{Bramwell}}, \citenamefont {{Christensen}}, \citenamefont {{Fortin}},
  \citenamefont {{Holdsworth}}, \citenamefont {{Jensen}}, \citenamefont
  {{Lise}}, \citenamefont {{L{\'o}pez}}, \citenamefont {{Nicodemi}},
  \citenamefont {{Pinton}},\ and\ \citenamefont {{Sellitto}}}]{BHP_2}%
  \BibitemOpen
  \bibfield  {author} {\bibinfo {author} {\bibfnamefont {S.~T.}\ \bibnamefont
  {{Bramwell}}}, \bibinfo {author} {\bibfnamefont {K.}~\bibnamefont
  {{Christensen}}}, \bibinfo {author} {\bibfnamefont {J.~Y.}\ \bibnamefont
  {{Fortin}}}, \bibinfo {author} {\bibfnamefont {P.~C.~W.}\ \bibnamefont
  {{Holdsworth}}}, \bibinfo {author} {\bibfnamefont {H.~J.}\ \bibnamefont
  {{Jensen}}}, \bibinfo {author} {\bibfnamefont {S.}~\bibnamefont {{Lise}}},
  \bibinfo {author} {\bibfnamefont {J.~M.}\ \bibnamefont {{L{\'o}pez}}},
  \bibinfo {author} {\bibfnamefont {M.}~\bibnamefont {{Nicodemi}}}, \bibinfo
  {author} {\bibfnamefont {J.~F.}\ \bibnamefont {{Pinton}}}, \ and\ \bibinfo
  {author} {\bibfnamefont {M.}~\bibnamefont {{Sellitto}}},\ }\href {\doibase
  10.1103/PhysRevLett.84.3744} {\bibfield  {journal} {\bibinfo  {journal}
  {\prl}\ }\textbf {\bibinfo {volume} {84}},\ \bibinfo {pages} {3744} (\bibinfo
  {year} {2000})}\BibitemShut {NoStop}%
\bibitem [{\citenamefont {{Portelli}}\ and\ \citenamefont
  {{Holdsworth}}(2002)}]{Portelli_02}%
  \BibitemOpen
  \bibfield  {author} {\bibinfo {author} {\bibfnamefont {B.}~\bibnamefont
  {{Portelli}}}\ and\ \bibinfo {author} {\bibfnamefont {P.~C.~W.}\ \bibnamefont
  {{Holdsworth}}},\ }\href {\doibase 10.1088/0305-4470/35/5/307} {\bibfield
  {journal} {\bibinfo  {journal} {Journal of Physics A Mathematical General}\
  }\textbf {\bibinfo {volume} {35}},\ \bibinfo {pages} {1231} (\bibinfo {year}
  {2002})}\BibitemShut {NoStop}%
\bibitem [{\citenamefont {Archambault}\ \emph {et~al.}(1997)\citenamefont
  {Archambault}, \citenamefont {Bramwell},\ and\ \citenamefont
  {Holdsworth}}]{Archambault_1997}%
  \BibitemOpen
  \bibfield  {author} {\bibinfo {author} {\bibfnamefont {P.}~\bibnamefont
  {Archambault}}, \bibinfo {author} {\bibfnamefont {S.~T.}\ \bibnamefont
  {Bramwell}}, \ and\ \bibinfo {author} {\bibfnamefont {P.~C.~W.}\ \bibnamefont
  {Holdsworth}},\ }\href {\doibase 10.1088/0305-4470/30/24/005} {\bibfield
  {journal} {\bibinfo  {journal} {J. Phys. A}\ }\textbf {\bibinfo {volume}
  {30}},\ \bibinfo {pages} {8363} (\bibinfo {year} {1997})}\BibitemShut
  {NoStop}%
\bibitem [{\citenamefont {Joubaud}\ \emph {et~al.}(2008)\citenamefont
  {Joubaud}, \citenamefont {Petrosyan}, \citenamefont {Ciliberto},\ and\
  \citenamefont {Garnier}}]{Joubaud2008}%
  \BibitemOpen
  \bibfield  {author} {\bibinfo {author} {\bibfnamefont {S.}~\bibnamefont
  {Joubaud}}, \bibinfo {author} {\bibfnamefont {A.}~\bibnamefont {Petrosyan}},
  \bibinfo {author} {\bibfnamefont {S.}~\bibnamefont {Ciliberto}}, \ and\
  \bibinfo {author} {\bibfnamefont {N.~B.}\ \bibnamefont {Garnier}},\ }\href
  {\doibase 10.1103/PhysRevLett.100.180601} {\bibfield  {journal} {\bibinfo
  {journal} {Phys. Rev. Lett.}\ }\textbf {\bibinfo {volume} {100}},\ \bibinfo
  {pages} {180601} (\bibinfo {year} {2008})}\BibitemShut {NoStop}%
\bibitem [{\citenamefont {Wilson}\ and\ \citenamefont
  {Kogut}(1974)}]{WilsonKogut_74}%
  \BibitemOpen
  \bibfield  {author} {\bibinfo {author} {\bibfnamefont {K.~G.}\ \bibnamefont
  {Wilson}}\ and\ \bibinfo {author} {\bibfnamefont {J.~B.}\ \bibnamefont
  {Kogut}},\ }\href {\doibase 10.1016/0370-1573(74)90023-4} {\bibfield
  {journal} {\bibinfo  {journal} {Phys. Rept.}\ }\textbf {\bibinfo {volume}
  {12}},\ \bibinfo {pages} {75} (\bibinfo {year} {1974})}\BibitemShut {NoStop}%
%%CITATION = PRPLC,12,75;%%
\bibitem [{\citenamefont {Bleher}\ and\ \citenamefont
  {Sinai}(1973)}]{Bleher1973}%
  \BibitemOpen
  \bibfield  {author} {\bibinfo {author} {\bibfnamefont {P.~M.}\ \bibnamefont
  {Bleher}}\ and\ \bibinfo {author} {\bibfnamefont {J.~G.}\ \bibnamefont
  {Sinai}},\ }\href {\doibase 10.1007/BF01645604} {\bibfield  {journal}
  {\bibinfo  {journal} {Communications in Mathematical Physics}\ }\textbf
  {\bibinfo {volume} {33}},\ \bibinfo {pages} {23} (\bibinfo {year}
  {1973})}\BibitemShut {NoStop}%
\bibitem [{\citenamefont {Bleher}\ and\ \citenamefont
  {Sinai}(1975)}]{Bleher1975}%
  \BibitemOpen
  \bibfield  {author} {\bibinfo {author} {\bibfnamefont {P.~M.}\ \bibnamefont
  {Bleher}}\ and\ \bibinfo {author} {\bibfnamefont {Y.~G.}\ \bibnamefont
  {Sinai}},\ }\href {\doibase 10.1007/BF01608331} {\bibfield  {journal}
  {\bibinfo  {journal} {Communications in Mathematical Physics}\ }\textbf
  {\bibinfo {volume} {45}},\ \bibinfo {pages} {247} (\bibinfo {year}
  {1975})}\BibitemShut {NoStop}%
\bibitem [{\citenamefont {Bleher}\ and\ \citenamefont
  {Major}(1987)}]{Bleher1987}%
  \BibitemOpen
  \bibfield  {author} {\bibinfo {author} {\bibfnamefont {P.~M.}\ \bibnamefont
  {Bleher}}\ and\ \bibinfo {author} {\bibfnamefont {P.}~\bibnamefont {Major}},\
  }\href {http://www.jstor.org/stable/2244058} {\bibfield  {journal} {\bibinfo
  {journal} {The Annals of Probability}\ }\textbf {\bibinfo {volume} {15}},\
  \bibinfo {pages} {431} (\bibinfo {year} {1987})}\BibitemShut {NoStop}%
\bibitem [{\citenamefont {Bleher}\ and\ \citenamefont
  {Major}(1988)}]{Bleher1988}%
  \BibitemOpen
  \bibfield  {author} {\bibinfo {author} {\bibfnamefont {P.~M.}\ \bibnamefont
  {Bleher}}\ and\ \bibinfo {author} {\bibfnamefont {P.}~\bibnamefont {Major}},\
  }\href {http://www.numdam.org/item/AIHPA_1988__49_1_7_0/} {\bibfield
  {journal} {\bibinfo  {journal} {Annales de l'I.H.P. Physique th\'eorique}\
  }\textbf {\bibinfo {volume} {49}},\ \bibinfo {pages} {7} (\bibinfo {year}
  {1988})}\BibitemShut {NoStop}%
\bibitem [{\citenamefont {Bleher}\ and\ \citenamefont
  {Major}(1989)}]{Bleher1989}%
  \BibitemOpen
  \bibfield  {author} {\bibinfo {author} {\bibfnamefont {P.~M.}\ \bibnamefont
  {Bleher}}\ and\ \bibinfo {author} {\bibfnamefont {P.}~\bibnamefont {Major}},\
  }\href@noop {} {\bibfield  {journal} {\bibinfo  {journal} {Communications in
  Mathematical Physics}\ }\textbf {\bibinfo {volume} {125}},\ \bibinfo {pages}
  {43 } (\bibinfo {year} {1989})}\BibitemShut {NoStop}%
\bibitem [{\citenamefont {Jona-Lasinio}(1975)}]{JonaLasino1975}%
  \BibitemOpen
  \bibfield  {author} {\bibinfo {author} {\bibfnamefont {G.}~\bibnamefont
  {Jona-Lasinio}},\ }\href@noop {} {\bibfield  {journal} {\bibinfo  {journal}
  {Il Nuovo Cimento B (1971-1996)}\ }\textbf {\bibinfo {volume} {26}},\
  \bibinfo {pages} {99} (\bibinfo {year} {1975})}\BibitemShut {NoStop}%
\bibitem [{\citenamefont {Cassandro}\ and\ \citenamefont
  {Jona-Lasinio}(1978)}]{Cassandro1978}%
  \BibitemOpen
  \bibfield  {author} {\bibinfo {author} {\bibfnamefont {M.}~\bibnamefont
  {Cassandro}}\ and\ \bibinfo {author} {\bibfnamefont {G.}~\bibnamefont
  {Jona-Lasinio}},\ }\href {\doibase 10.1080/00018737800101504} {\bibfield
  {journal} {\bibinfo  {journal} {Advances in Physics}\ }\textbf {\bibinfo
  {volume} {27}},\ \bibinfo {pages} {913} (\bibinfo {year} {1978})}\BibitemShut
  {NoStop}%
\bibitem [{\citenamefont {Jona-Lasinio}(2001)}]{JonaLasino2001}%
  \BibitemOpen
  \bibfield  {author} {\bibinfo {author} {\bibfnamefont {G.}~\bibnamefont
  {Jona-Lasinio}},\ }\href {\doibase
  https://doi.org/10.1016/S0370-1573(01)00042-4} {\bibfield  {journal}
  {\bibinfo  {journal} {Physics Reports}\ }\textbf {\bibinfo {volume} {352}},\
  \bibinfo {pages} {439} (\bibinfo {year} {2001})}\BibitemShut {NoStop}%
\bibitem [{\citenamefont {Bruce}\ \emph {et~al.}(1979)\citenamefont {Bruce},
  \citenamefont {Schneider},\ and\ \citenamefont {Stoll}}]{Bruce1979}%
  \BibitemOpen
  \bibfield  {author} {\bibinfo {author} {\bibfnamefont {A.~D.}\ \bibnamefont
  {Bruce}}, \bibinfo {author} {\bibfnamefont {T.}~\bibnamefont {Schneider}}, \
  and\ \bibinfo {author} {\bibfnamefont {E.}~\bibnamefont {Stoll}},\ }\href
  {\doibase 10.1103/PhysRevLett.43.1284} {\bibfield  {journal} {\bibinfo
  {journal} {Phys. Rev. Lett.}\ }\textbf {\bibinfo {volume} {43}},\ \bibinfo
  {pages} {1284} (\bibinfo {year} {1979})}\BibitemShut {NoStop}%
\bibitem [{\citenamefont {Rudnick}\ \emph {et~al.}(1985)\citenamefont
  {Rudnick}, \citenamefont {Guo},\ and\ \citenamefont {Jasnow}}]{Rudnick1985}%
  \BibitemOpen
  \bibfield  {author} {\bibinfo {author} {\bibfnamefont {J.}~\bibnamefont
  {Rudnick}}, \bibinfo {author} {\bibfnamefont {H.}~\bibnamefont {Guo}}, \ and\
  \bibinfo {author} {\bibfnamefont {D.}~\bibnamefont {Jasnow}},\ }\href
  {\doibase 10.1007/BF01009013} {\bibfield  {journal} {\bibinfo  {journal} {J.
  Stat. Phys.}\ }\textbf {\bibinfo {volume} {41}},\ \bibinfo {pages} {353}
  (\bibinfo {year} {1985})}\BibitemShut {NoStop}%
\bibitem [{\citenamefont {Eisenriegler}\ and\ \citenamefont
  {Tomaschitz}(1987)}]{Eisenriegler1987}%
  \BibitemOpen
  \bibfield  {author} {\bibinfo {author} {\bibfnamefont {E.}~\bibnamefont
  {Eisenriegler}}\ and\ \bibinfo {author} {\bibfnamefont {R.}~\bibnamefont
  {Tomaschitz}},\ }\href {\doibase 10.1103/PhysRevB.35.4876} {\bibfield
  {journal} {\bibinfo  {journal} {Phys. Rev. B}\ }\textbf {\bibinfo {volume}
  {35}},\ \bibinfo {pages} {4876} (\bibinfo {year} {1987})}\BibitemShut
  {NoStop}%
\bibitem [{\citenamefont {Hilfer}(1993)}]{Hilfer1993}%
  \BibitemOpen
  \bibfield  {author} {\bibinfo {author} {\bibfnamefont {R.}~\bibnamefont
  {Hilfer}},\ }\href {\doibase 10.1142/S0217979293003711} {\bibfield  {journal}
  {\bibinfo  {journal} {International Journal of Modern Physics B}\ }\textbf
  {\bibinfo {volume} {07}},\ \bibinfo {pages} {4371} (\bibinfo {year}
  {1993})}\BibitemShut {NoStop}%
\bibitem [{\citenamefont {Hilfer}\ and\ \citenamefont
  {Wilding}(1995)}]{Hilfer1995}%
  \BibitemOpen
  \bibfield  {author} {\bibinfo {author} {\bibfnamefont {R.}~\bibnamefont
  {Hilfer}}\ and\ \bibinfo {author} {\bibfnamefont {N.~B.}\ \bibnamefont
  {Wilding}},\ }\href {\doibase 10.1088/0305-4470/28/10/001} {\bibfield
  {journal} {\bibinfo  {journal} {Journal of Physics A: Mathematical and
  General}\ }\textbf {\bibinfo {volume} {28}},\ \bibinfo {pages} {L281}
  (\bibinfo {year} {1995})}\BibitemShut {NoStop}%
\bibitem [{\citenamefont {Esser}\ \emph {et~al.}(1995)\citenamefont {Esser},
  \citenamefont {Dohm},\ and\ \citenamefont {Chen}}]{Esser1995}%
  \BibitemOpen
  \bibfield  {author} {\bibinfo {author} {\bibfnamefont {A.}~\bibnamefont
  {Esser}}, \bibinfo {author} {\bibfnamefont {V.}~\bibnamefont {Dohm}}, \ and\
  \bibinfo {author} {\bibfnamefont {X.}~\bibnamefont {Chen}},\ }\href {\doibase
  https://doi.org/10.1016/0378-4371(95)00264-2} {\bibfield  {journal} {\bibinfo
   {journal} {Physica A: Statistical Mechanics and its Applications}\ }\textbf
  {\bibinfo {volume} {222}},\ \bibinfo {pages} {355} (\bibinfo {year}
  {1995})}\BibitemShut {NoStop}%
\bibitem [{\citenamefont {Chen}\ \emph {et~al.}(1996)\citenamefont {Chen},
  \citenamefont {Dohm},\ and\ \citenamefont {Schultka}}]{ChenDohm_96}%
  \BibitemOpen
  \bibfield  {author} {\bibinfo {author} {\bibfnamefont {X.~S.}\ \bibnamefont
  {Chen}}, \bibinfo {author} {\bibfnamefont {V.}~\bibnamefont {Dohm}}, \ and\
  \bibinfo {author} {\bibfnamefont {N.}~\bibnamefont {Schultka}},\ }\href
  {\doibase 10.1103/PhysRevLett.77.3641} {\bibfield  {journal} {\bibinfo
  {journal} {Phys. Rev. Lett.}\ }\textbf {\bibinfo {volume} {77}},\ \bibinfo
  {pages} {3641} (\bibinfo {year} {1996})}\BibitemShut {NoStop}%
\bibitem [{\citenamefont {Chen}\ and\ \citenamefont {Dohm}(1998)}]{Chen1998}%
  \BibitemOpen
  \bibfield  {author} {\bibinfo {author} {\bibfnamefont {X.~S.}\ \bibnamefont
  {Chen}}\ and\ \bibinfo {author} {\bibfnamefont {V.}~\bibnamefont {Dohm}},\
  }\href {\doibase 10.1142/S0217979298000703} {\bibfield  {journal} {\bibinfo
  {journal} {International Journal of Modern Physics B}\ }\textbf {\bibinfo
  {volume} {12}},\ \bibinfo {pages} {1277} (\bibinfo {year}
  {1998})}\BibitemShut {NoStop}%
\bibitem [{\citenamefont {Rudnick}\ \emph {et~al.}(1998)\citenamefont
  {Rudnick}, \citenamefont {Lay},\ and\ \citenamefont {Jasnow}}]{Rudnick1998}%
  \BibitemOpen
  \bibfield  {author} {\bibinfo {author} {\bibfnamefont {J.}~\bibnamefont
  {Rudnick}}, \bibinfo {author} {\bibfnamefont {W.}~\bibnamefont {Lay}}, \ and\
  \bibinfo {author} {\bibfnamefont {D.}~\bibnamefont {Jasnow}},\ }\href
  {\doibase 10.1103/PhysRevE.58.2902} {\bibfield  {journal} {\bibinfo
  {journal} {Phys. Rev. E}\ }\textbf {\bibinfo {volume} {58}},\ \bibinfo
  {pages} {2902} (\bibinfo {year} {1998})}\BibitemShut {NoStop}%
\bibitem [{\citenamefont {Sahu}\ \emph {et~al.}(2024)\citenamefont {Sahu},
  \citenamefont {Delamotte},\ and\ \citenamefont {Rançon}}]{Sahu2024}%
  \BibitemOpen
  \bibfield  {author} {\bibinfo {author} {\bibfnamefont {S.}~\bibnamefont
  {Sahu}}, \bibinfo {author} {\bibfnamefont {B.}~\bibnamefont {Delamotte}}, \
  and\ \bibinfo {author} {\bibfnamefont {A.}~\bibnamefont {Rançon}},\ }\href
  {https://arxiv.org/abs/2407.12603} {} (\bibinfo {year} {2024}),\ \Eprint
  {http://arxiv.org/abs/2407.12603} {arXiv:2407.12603} \BibitemShut {NoStop}%
\bibitem [{\citenamefont {Berges}\ \emph {et~al.}(2002)\citenamefont {Berges},
  \citenamefont {Tetradis},\ and\ \citenamefont {Wetterich}}]{Berges_02}%
  \BibitemOpen
  \bibfield  {author} {\bibinfo {author} {\bibfnamefont {J.}~\bibnamefont
  {Berges}}, \bibinfo {author} {\bibfnamefont {N.}~\bibnamefont {Tetradis}}, \
  and\ \bibinfo {author} {\bibfnamefont {C.}~\bibnamefont {Wetterich}},\ }\href
  {\doibase DOI: 10.1016/S0370-1573(01)00098-9} {\bibfield  {journal} {\bibinfo
   {journal} {Physics Reports}\ }\textbf {\bibinfo {volume} {363}},\ \bibinfo
  {pages} {223 } (\bibinfo {year} {2002})}\BibitemShut {NoStop}%
\bibitem [{\citenamefont {Dupuis}\ \emph {et~al.}(2021)\citenamefont {Dupuis},
  \citenamefont {Canet}, \citenamefont {Eichhorn}, \citenamefont {Metzner},
  \citenamefont {Pawlowski}, \citenamefont {Tissier},\ and\ \citenamefont
  {Wschebor}}]{Dupuis2021}%
  \BibitemOpen
  \bibfield  {author} {\bibinfo {author} {\bibfnamefont {N.}~\bibnamefont
  {Dupuis}}, \bibinfo {author} {\bibfnamefont {L.}~\bibnamefont {Canet}},
  \bibinfo {author} {\bibfnamefont {A.}~\bibnamefont {Eichhorn}}, \bibinfo
  {author} {\bibfnamefont {W.}~\bibnamefont {Metzner}}, \bibinfo {author}
  {\bibfnamefont {J.}~\bibnamefont {Pawlowski}}, \bibinfo {author}
  {\bibfnamefont {M.}~\bibnamefont {Tissier}}, \ and\ \bibinfo {author}
  {\bibfnamefont {N.}~\bibnamefont {Wschebor}},\ }\href {\doibase
  https://doi.org/10.1016/j.physrep.2021.01.001} {\bibfield  {journal}
  {\bibinfo  {journal} {Physics Reports}\ }\textbf {\bibinfo {volume} {910}},\
  \bibinfo {pages} {1} (\bibinfo {year} {2021})}\BibitemShut {NoStop}%
\bibitem [{\citenamefont {Moshe}\ and\ \citenamefont
  {Zinn-Justin}(2003)}]{Moshe2003}%
  \BibitemOpen
  \bibfield  {author} {\bibinfo {author} {\bibfnamefont {M.}~\bibnamefont
  {Moshe}}\ and\ \bibinfo {author} {\bibfnamefont {J.}~\bibnamefont
  {Zinn-Justin}},\ }\href {\doibase
  https://doi.org/10.1016/S0370-1573(03)00263-1} {\bibfield  {journal}
  {\bibinfo  {journal} {Physics Reports}\ }\textbf {\bibinfo {volume} {385}},\
  \bibinfo {pages} {69 } (\bibinfo {year} {2003})}\BibitemShut {NoStop}%
\bibitem [{\citenamefont {Balog}\ \emph {et~al.}(2022)\citenamefont {Balog},
  \citenamefont {Ran\ifmmode~\mbox{\c{c}}\else \c{c}\fi{}on},\ and\
  \citenamefont {Delamotte}}]{Balog22}%
  \BibitemOpen
  \bibfield  {author} {\bibinfo {author} {\bibfnamefont {I.}~\bibnamefont
  {Balog}}, \bibinfo {author} {\bibfnamefont {A.}~\bibnamefont
  {Ran\ifmmode~\mbox{\c{c}}\else \c{c}\fi{}on}}, \ and\ \bibinfo {author}
  {\bibfnamefont {B.}~\bibnamefont {Delamotte}},\ }\href {\doibase
  10.1103/PhysRevLett.129.210602} {\bibfield  {journal} {\bibinfo  {journal}
  {Phys. Rev. Lett.}\ }\textbf {\bibinfo {volume} {129}},\ \bibinfo {pages}
  {210602} (\bibinfo {year} {2022})}\BibitemShut {NoStop}%
\bibitem [{\citenamefont {Balog}\ \emph {et~al.}(2024)\citenamefont {Balog},
  \citenamefont {Delamotte},\ and\ \citenamefont {Rançon}}]{Balog2024}%
  \BibitemOpen
  \bibfield  {author} {\bibinfo {author} {\bibfnamefont {I.}~\bibnamefont
  {Balog}}, \bibinfo {author} {\bibfnamefont {B.}~\bibnamefont {Delamotte}}, \
  and\ \bibinfo {author} {\bibfnamefont {A.}~\bibnamefont {Rançon}},\ }\href
  {https://arxiv.org/abs/2409.01250} {\enquote {\bibinfo {title} {Universal and
  non-universal large deviations in critical systems},}\ } (\bibinfo {year}
  {2024}),\ \Eprint {http://arxiv.org/abs/2409.01250} {arXiv:2409.01250}
  \BibitemShut {NoStop}%
\bibitem [{\citenamefont {Pel\'aez}\ and\ \citenamefont
  {Wschebor}(2016)}]{pelaez2016}%
  \BibitemOpen
  \bibfield  {author} {\bibinfo {author} {\bibfnamefont {M.}~\bibnamefont
  {Pel\'aez}}\ and\ \bibinfo {author} {\bibfnamefont {N.}~\bibnamefont
  {Wschebor}},\ }\href {\doibase 10.1103/PhysRevE.94.042136} {\bibfield
  {journal} {\bibinfo  {journal} {Phys. Rev. E}\ }\textbf {\bibinfo {volume}
  {94}},\ \bibinfo {pages} {042136} (\bibinfo {year} {2016})}\BibitemShut
  {NoStop}%
\bibitem [{\citenamefont {Wetterich}(1993{\natexlab{a}})}]{Wetterich1993}%
  \BibitemOpen
  \bibfield  {author} {\bibinfo {author} {\bibfnamefont {C.}~\bibnamefont
  {Wetterich}},\ }\href {\doibase DOI: 10.1016/0370-2693(93)90726-X} {\bibfield
   {journal} {\bibinfo  {journal} {Physics Letters B}\ }\textbf {\bibinfo
  {volume} {301}},\ \bibinfo {pages} {90 } (\bibinfo {year}
  {1993}{\natexlab{a}})}\BibitemShut {NoStop}%
\bibitem [{\citenamefont {Wetterich}(1993{\natexlab{b}})}]{Wetterich1993a}%
  \BibitemOpen
  \bibfield  {author} {\bibinfo {author} {\bibfnamefont {C.}~\bibnamefont
  {Wetterich}},\ }\href {http://dx.doi.org/10.1007/BF01560044} {\bibfield
  {journal} {\bibinfo  {journal} {Zeitschrift f\"ur Physik C}\ }\textbf
  {\bibinfo {volume} {60}},\ \bibinfo {pages} {461} (\bibinfo {year}
  {1993}{\natexlab{b}})}\BibitemShut {NoStop}%
\bibitem [{\citenamefont {Wetterich}(1993{\natexlab{c}})}]{Wetterich1993b}%
  \BibitemOpen
  \bibfield  {author} {\bibinfo {author} {\bibfnamefont {C.}~\bibnamefont
  {Wetterich}},\ }\href {http://dx.doi.org/10.1007/BF01474340} {\bibfield
  {journal} {\bibinfo  {journal} {Zeitschrift f\"ur Physik C Particles and
  Fields}\ }\textbf {\bibinfo {volume} {57}},\ \bibinfo {pages} {451} (\bibinfo
  {year} {1993}{\natexlab{c}})}\BibitemShut {NoStop}%
\bibitem [{\citenamefont {Fister}\ and\ \citenamefont
  {Pawlowski}(2015)}]{Fister2015}%
  \BibitemOpen
  \bibfield  {author} {\bibinfo {author} {\bibfnamefont {L.}~\bibnamefont
  {Fister}}\ and\ \bibinfo {author} {\bibfnamefont {J.~M.}\ \bibnamefont
  {Pawlowski}},\ }\href {\doibase 10.1103/PhysRevD.92.076009} {\bibfield
  {journal} {\bibinfo  {journal} {Phys. Rev. D}\ }\textbf {\bibinfo {volume}
  {92}},\ \bibinfo {pages} {076009} (\bibinfo {year} {2015})}\BibitemShut
  {NoStop}%
\bibitem [{\citenamefont {Balog}\ \emph {et~al.}(2019)\citenamefont {Balog},
  \citenamefont {Chat{\'e}}, \citenamefont {Delamotte}, \citenamefont
  {Marohni{\'c}},\ and\ \citenamefont {Wschebor}}]{Balog2019}%
  \BibitemOpen
  \bibfield  {author} {\bibinfo {author} {\bibfnamefont {I.}~\bibnamefont
  {Balog}}, \bibinfo {author} {\bibfnamefont {H.}~\bibnamefont {Chat{\'e}}},
  \bibinfo {author} {\bibfnamefont {B.}~\bibnamefont {Delamotte}}, \bibinfo
  {author} {\bibfnamefont {M.}~\bibnamefont {Marohni{\'c}}}, \ and\ \bibinfo
  {author} {\bibfnamefont {N.}~\bibnamefont {Wschebor}},\ }\href {\doibase
  10.1103/PhysRevLett.123.240604} {\bibfield  {journal} {\bibinfo  {journal}
  {Physical Review Letters}\ }\textbf {\bibinfo {volume} {123}},\ \bibinfo
  {pages} {240604} (\bibinfo {year} {2019})}\BibitemShut {NoStop}%
\bibitem [{\citenamefont {De~Polsi}\ and\ \citenamefont
  {Wschebor}(2022)}]{DePolsi2022}%
  \BibitemOpen
  \bibfield  {author} {\bibinfo {author} {\bibfnamefont {G.}~\bibnamefont
  {De~Polsi}}\ and\ \bibinfo {author} {\bibfnamefont {N.}~\bibnamefont
  {Wschebor}},\ }\href {\doibase 10.1103/PhysRevE.106.024111} {\bibfield
  {journal} {\bibinfo  {journal} {Phys. Rev. E}\ }\textbf {\bibinfo {volume}
  {106}},\ \bibinfo {pages} {024111} (\bibinfo {year} {2022})}\BibitemShut
  {NoStop}%
\bibitem [{\citenamefont {De~Polsi}\ \emph {et~al.}(2020)\citenamefont
  {De~Polsi}, \citenamefont {Balog}, \citenamefont {Tissier},\ and\
  \citenamefont {Wschebor}}]{depolsi2020}%
  \BibitemOpen
  \bibfield  {author} {\bibinfo {author} {\bibfnamefont {G.}~\bibnamefont
  {De~Polsi}}, \bibinfo {author} {\bibfnamefont {I.}~\bibnamefont {Balog}},
  \bibinfo {author} {\bibfnamefont {M.}~\bibnamefont {Tissier}}, \ and\
  \bibinfo {author} {\bibfnamefont {N.}~\bibnamefont {Wschebor}},\ }\href
  {\doibase 10.1103/PhysRevE.101.042113} {\bibfield  {journal} {\bibinfo
  {journal} {Phys. Rev. E}\ }\textbf {\bibinfo {volume} {101}},\ \bibinfo
  {pages} {042113} (\bibinfo {year} {2020})}\BibitemShut {NoStop}%
\bibitem [{Ros()}]{Rose2031}%
  \BibitemOpen
  \href@noop {} {}\bibinfo {note} {F. Rose, I. Balog, A. Ran\c{c}on, to be
  published.}\BibitemShut {Stop}%
\bibitem [{\citenamefont {Tetradis}\ and\ \citenamefont
  {Litim}(1996)}]{Tetradis1996}%
  \BibitemOpen
  \bibfield  {author} {\bibinfo {author} {\bibfnamefont {N.}~\bibnamefont
  {Tetradis}}\ and\ \bibinfo {author} {\bibfnamefont {D.}~\bibnamefont
  {Litim}},\ }\href {\doibase https://doi.org/10.1016/0550-3213(95)00642-7}
  {\bibfield  {journal} {\bibinfo  {journal} {Nuclear Physics B}\ }\textbf
  {\bibinfo {volume} {464}},\ \bibinfo {pages} {492} (\bibinfo {year}
  {1996})}\BibitemShut {NoStop}%
\bibitem [{\citenamefont {Knorr}(2021)}]{knorr2021}%
  \BibitemOpen
  \bibfield  {author} {\bibinfo {author} {\bibfnamefont {B.}~\bibnamefont
  {Knorr}},\ }\href {\doibase 10.1088/1751-8121/ac00d4} {\bibfield  {journal}
  {\bibinfo  {journal} {Journal of Physics A: Mathematical and Theoretical}\
  }\textbf {\bibinfo {volume} {54}},\ \bibinfo {pages} {275401} (\bibinfo
  {year} {2021})}\BibitemShut {NoStop}%
\bibitem [{\citenamefont {Wolff}(1989)}]{Wolff1989}%
  \BibitemOpen
  \bibfield  {author} {\bibinfo {author} {\bibfnamefont {U.}~\bibnamefont
  {Wolff}},\ }\href {\doibase 10.1103/PhysRevLett.62.361} {\bibfield  {journal}
  {\bibinfo  {journal} {Phys. Rev. Lett.}\ }\textbf {\bibinfo {volume} {62}},\
  \bibinfo {pages} {361} (\bibinfo {year} {1989})}\BibitemShut {NoStop}%
\bibitem [{\citenamefont {Newman}\ and\ \citenamefont
  {Barkema}(1999)}]{newman1999monte}%
  \BibitemOpen
  \bibfield  {author} {\bibinfo {author} {\bibfnamefont {M.}~\bibnamefont
  {Newman}}\ and\ \bibinfo {author} {\bibfnamefont {G.}~\bibnamefont
  {Barkema}},\ }\href {https://books.google.hr/books?id=KKL2nQEACAAJ} {\emph
  {\bibinfo {title} {Monte Carlo Methods in Statistical Physics}}}\ (\bibinfo
  {publisher} {Clarendon Press},\ \bibinfo {year} {1999})\BibitemShut {NoStop}%
\bibitem [{\citenamefont {Butera}\ and\ \citenamefont
  {Comi}(1998)}]{butera1998}%
  \BibitemOpen
  \bibfield  {author} {\bibinfo {author} {\bibfnamefont {P.}~\bibnamefont
  {Butera}}\ and\ \bibinfo {author} {\bibfnamefont {M.}~\bibnamefont {Comi}},\
  }\href {\doibase 10.1103/PhysRevB.58.11552} {\bibfield  {journal} {\bibinfo
  {journal} {Phys. Rev. B}\ }\textbf {\bibinfo {volume} {58}},\ \bibinfo
  {pages} {11552} (\bibinfo {year} {1998})}\BibitemShut {NoStop}%
\bibitem [{\citenamefont {Hasenbusch}(2008)}]{hasenbusch2008}%
  \BibitemOpen
  \bibfield  {author} {\bibinfo {author} {\bibfnamefont {M.}~\bibnamefont
  {Hasenbusch}},\ }\href {\doibase 10.1088/1742-5468/2008/12/P12006} {\bibfield
   {journal} {\bibinfo  {journal} {Journal of Statistical Mechanics: Theory and
  Experiment}\ }\textbf {\bibinfo {volume} {2008}},\ \bibinfo {pages} {P12006}
  (\bibinfo {year} {2008})}\BibitemShut {NoStop}%
\bibitem [{\citenamefont {Deng}\ \emph {et~al.}(2005)\citenamefont {Deng},
  \citenamefont {Bl\"ote},\ and\ \citenamefont {Nightingale}}]{Deng2005}%
  \BibitemOpen
  \bibfield  {author} {\bibinfo {author} {\bibfnamefont {Y.}~\bibnamefont
  {Deng}}, \bibinfo {author} {\bibfnamefont {H.~W.~J.}\ \bibnamefont
  {Bl\"ote}}, \ and\ \bibinfo {author} {\bibfnamefont {M.~P.}\ \bibnamefont
  {Nightingale}},\ }\href {\doibase 10.1103/PhysRevE.72.016128} {\bibfield
  {journal} {\bibinfo  {journal} {Phys. Rev. E}\ }\textbf {\bibinfo {volume}
  {72}},\ \bibinfo {pages} {016128} (\bibinfo {year} {2005})}\BibitemShut
  {NoStop}%
\bibitem [{\citenamefont {Kanaya}\ and\ \citenamefont
  {Kaya}(1995)}]{Kanaya1995}%
  \BibitemOpen
  \bibfield  {author} {\bibinfo {author} {\bibfnamefont {K.}~\bibnamefont
  {Kanaya}}\ and\ \bibinfo {author} {\bibfnamefont {S.}~\bibnamefont {Kaya}},\
  }\href {\doibase 10.1103/PhysRevD.51.2404} {\bibfield  {journal} {\bibinfo
  {journal} {Phys. Rev. D}\ }\textbf {\bibinfo {volume} {51}},\ \bibinfo
  {pages} {2404} (\bibinfo {year} {1995})}\BibitemShut {NoStop}%
\end{thebibliography}%
\bibliographystyle{apsrev4-1} % Tell bibtex which bibliography style to use

\end{document}